\documentclass[twocolumn,aps,floatfix,prc]{revtex4-1}
\usepackage{graphicx}
\usepackage{dcolumn}
\usepackage{bm}
\usepackage{rotating}
\usepackage{longtable}
\usepackage{float}
\usepackage{eucal}
\usepackage{csquotes}
\usepackage{xcolor}
\usepackage{amsmath}
\usepackage{amssymb}
\usepackage{natbib}
\usepackage[T1]{fontenc}
\usepackage[pagewise]{lineno}
\usepackage{amstext}
\usepackage{ulem}

\DeclareUnicodeCharacter{2212}{-}
\begin{document}

\title{\centering {Shape evolution in the rapidly rotating $^{140}$Gd nucleus}}

\author{H. Pai$^{1}$}
\email{haridas.pai@eli-np.ro; hari.vecc@gmail.com}
\author{S. Rajbanshi$^{2}$}
\author{Somnath Nag$^{3}$}
\author{Sajad Ali$^{4}$}
\author{R. Palit$^{5}$}
\author{G. Mukherjee$^{6,7}$}
\author{F. S. Babra$^{5}$}
\author{R. Banik$^{8}$}
\author{Soumik Bhattacharya$^{6,7}$}
\author{S. Biswas$^{5}$}
\altaffiliation{Paul Scherrer Institute, Forschungsstrasse 111, 5232, Villigen PSI, Switzerland}
\author{S. Chakraborty$^{6}$}
\author{R. Donthi$^{5}$}
\author{S. Jadhav$^{5}$}
\author{Md. S. R. Laskar$^{5}$}
\author{B. S. Naidu$^{5}$}
\author{S. Nandi$^{6,7}$}
\altaffiliation{Present address: Argonne National Laboratory, Argonne, Illinois 60439, USA}
\author{A. Goswami$^{9}$}
\affiliation{$^{1}$Extreme Light Infrastructure - Nuclear Physics, Horia Hulubei National Institute for R$\&$D in Physics and Nuclear Engineering, Bucharest-Magurele, 077125, Romania}
\affiliation{$^{2}$Department of Physics, Presidency University, Kolkata 700073, India}
\affiliation{$^{3}$Department of Physics, Indian Institute of Technology (BHU), Varanasi 221005, India}
\affiliation{$^{4}$Government General Degree College at Pedong, Kalimpong, 734311, India}
\affiliation{$^5$Tata Institute of Fundamental Research, Mumbai 400005, India}
\affiliation{$^6$Variable Energy Cyclotron Centre,1/AF, Bidhannagar, Kolkata 700064, India}
\affiliation{$^7$Homi Bhabha National Institute, Training School Complex, Anushakti Nagar, Mumnai 400094, India}
\affiliation{$^8$Institute of Engineering and Management, Saltlake Sector V, Kolkata 700091, India}
\affiliation{$^9$Saha Institute of Nuclear Physics, 1/AF, Bidhannagar, Kolkata 700064, India}

\date{\today}

\begin{abstract}
Ground state band of $^{140}$Gd has been investigated following their population in the $^{112}$Sn($^{35}$Cl,~$\alpha$p2n)$^{140}$Gd reaction at 195 MeV of beam energy using a large array of Compton suppressed HPGe clovers as the detection setup. Apart from other spectroscopic measurements, level lifetimes of the states have been extracted using the Doppler Shift Attenuation Method. Extracted quadrupole moment along with the pairing independent cranked Nilsson-Strutinsky model calculations for the quadrupole band reveal that the nucleus preferably attains triaxiality with $\gamma$ = -30$^\circ$. The calculation though shows a slight possibility of rotation around the longest possible principal axis at high spin $\sim$ 30$\hbar$ which is beyond the scope of the present experiment.

\end{abstract}

\pacs{21.10.Re, 21.10.Tg, 21.60.Ev, 23.20.Lv, 27.60.+j}

\maketitle

\section{Introduction}
The structure and shape of the nucleus have been intriguingly studied over the long decades. In order to describe the nuclear fission process,  Bohr and Wheeler assumed for the first time that a nucleus can be regarded as a drop of liquid \cite{bohandwh}. Development of the shell model hinted that the accumulation of p-n (proton-neutron) interaction strength will lead to additional configuration mixing and deviation from the spherical symmetry even in the ground state. This produces the asymmetry in mass distribution which results in a permanent stable deformed shape, oblate or prolate, in nuclei. Within the perspective of quantum phenomenon, collective rotation is then only possible around an axis perpendicular to the symmetry axis. Accumulation of the p-n interaction strength becomes largest for the mid-shell nuclei consequently regular sequences characterized by E2 intraband transitions have been observed in their level structure \cite{abohr, defor1, defor2}. However, the rotation around any of the three principal axes; short, long, and intermediate axes, is possible for the nuclei having triaxial nuclear mass distribution \cite{pmoller} and triaxiality plays an important role in rotating nuclei. The energy of the triaxial mass distribution depends on the moment of inertia which is characterized by the deformation parameters $\varepsilon_2$ and $\gamma$. The energy of rotation around the shortest principal axis becomes favourable for $\gamma$ $\approx$ 30$^{\circ}$ which is observed for nuclei in $A$ $\sim$ 60, 110, 140 and 160 mass regions \cite{avafan, gbhagm, cmpetra}. Rotation around the intermediate axis becomes energetically favourable for $\gamma$ $\approx$ $-$30$^{\circ}$ and when pairing correlations are considered \cite{rbengts}, as evident from the recent studies in $^{77}$Kr \cite{tseinha}. Furthermore classically unfavoured rotation about the longest principal axis {corresponds to $\gamma$ $\approx$ $-$75$^{\circ}$ as observed in $^{142}$Gd \cite{bgcarls}}, $^{137, 138}$Nd \cite{cmpetra11, cmpetr222}, $^{193}$Tl \cite{jndayis11} and Au isotopes \cite{egueoru11}. Though these three types of rotation are observed in different nuclei, their existence in a single nucleus has been predicted only in $^{138}$Nd \cite{cmpetra11} on the theoretical basis of the tilted axis cranking model (TAC) and cranked shell model (CSM) and the cranked Nilsson + BCS formalism without any transition probability (B(M1) and B(E2)) measurements, thereby the proposition was tentative only.

The $\gamma$-soft nuclei in the $A$ $\sim$ 140 mass region are crucial, in the present perspective, due to their favoured triaxial mass distribution. For this reason, the rotation around three different axes would appear in their level structure \cite{cmpetr1, cmpetr2}. In this region, the nuclei in the vicinity of the shell closures are weakly deformed at low and intermediate spin, evolving to a well-deformed shape with increasing spin/excitation energy \cite{rbroda1, nredon1, jpeders, cmpetr1, cmpetr2, cmpetr111, cmpetr222}.

Indeed, recent results in $^{142}$Gd nucleus highlight the existence of long-axis rotation which is preferred at high spin only due to alignment of extra neutron holes. This agreement was consistent only for one transition (26$^{+}$ $\rightarrow$ 24$^{+}$) as far as the measured B(E2) transition probability is concerned that were not able to follow the evolution of the structure. It is then of special interest to search for cases with rotation around the classically unfavoured longest principal axis for $^{140}$Gd as it has two extra neutron holes w.r.t the $^{142}$Gd nucleus. Thus long axis rotation may be possible at lower spin. With this motivation, the quadrupole moment i.e. B(E2) rates of the states in $^{140}$Gd have been measured and compared with the CNS calculations. The results show the nucleus has intermediate axis rotation (gamma $\sim$ -25$^o$) at lower spin which evolves to the long axis (gamma ~ $-$75$^o$) at higher spin (after crossing $I$ $\sim$ 12). In $^{140}$Gd, the ground-state band was well established without the level lifetimes information~\cite{nnd1,nnd2,nnd3}.


\section{EXPERIMENTAL DETAILS AND RESULTS}

The gamma-ray spectroscopy of $^{140}$Gd has been studied at 14-UD BARC-TIFR Pelletron at Mumbai, India using the eleven Compton-suppressed HPGe clover detectors at the time of the experiment.
The excited states of $^{140}$Gd 
were populated by fusion evaporation reactions $^{112}$Sn($^{35}$Cl,~$\alpha$p2n)$^{140}$Gd at 195 MeV of beam energy.
Target was 2.44 mg/cm$^2$ thick $^{112}$Sn (99.6\% enriched) with a 8.8 mg/cm$^2$ $^{208}$Pb backing~\cite{112Sn}.
The clover detectors were arranged in four angles with three clovers each at -23$^\circ$, -40$^\circ$ and, 90$^\circ$  angles while two clovers were at -65$^\circ$ angle. The digital data acquisition system
was used to record the trigger less, time-stamped data~\cite{dsp}.
About 9.6$\times$10$^{8}$ such coincidence events were recorded.
The data were sorted into different symmetric and angle dependent $E_{\gamma}$ - $E_{\gamma}$ matrices and, $E_{\gamma}$-$E_{\gamma}$-$E_{\gamma}$ cube with the help of the Multi pARameter time-stamped based COincidence Search program (MARCOS)~\cite{dsp,194tl,pah} and analyzed using the INGASORT and the RADWARE packages~\cite{rb, radford1, radford2}. 
The clover detectors were calibrated for $\gamma$-ray energies and efficiencies by using the $^{133}$Ba and $^{152}$Eu radioactive sources. 

\begin{figure}[t]
\begin{center}
\includegraphics[clip=true,width=7.8cm,angle=0]{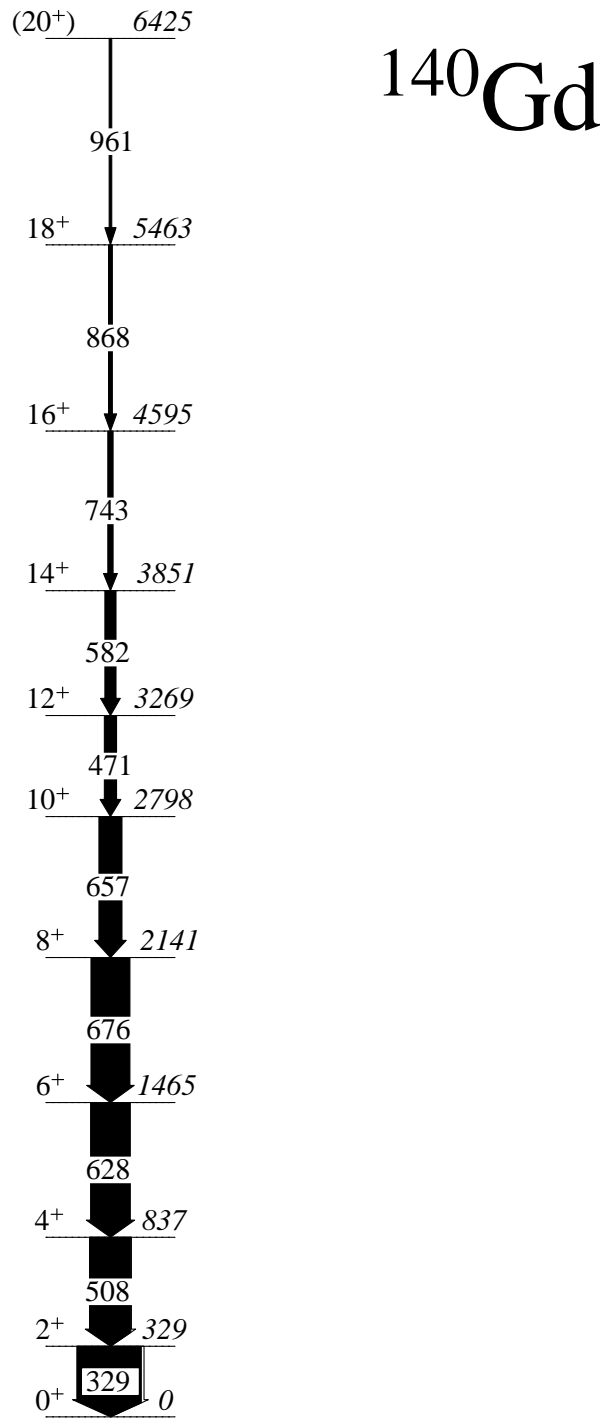} 
\caption{Partial level scheme of $^{140}$Gd (ground state band (QB I)), deduced from this work. }
\label{LS}
\end{center}
\end{figure}

\begin{table*}
\caption{\label{lifetime} The ADO ratio ($R_{ADO}$), polarization asymmetry ($\Delta_{\text{asym}}$), level lifetimes ($\tau$), and the corresponding $B(E2)$ transitions rates, and the measured transition quadrupole moments ($Q_{t}$) for the $\gamma$ transitions of the yrast band in $^{140}$Gd.}
\begin{ruledtabular}
\begin{tabular}{ccccccccccc}


$J_{i}$      &$J_{f}$    & $E_\gamma$\footnotemark[1] & Assign. & $I_{\gamma}$\footnotemark[2] & $R_{ADO}$ & $\Delta_{\text{asym}}$ & $\tau$\footnotemark[3]   &  $B(E2)$ &  $Q_{t}$ \\

[ $\hbar$ ]  & [ $\hbar$ ]& [ keV ]    &  &                  &                 &          &   [ ps ]  & [ $e^2b^2$ ] &[ $eb$ ]\\

\hline\hline

2$^{+}$ & 0$^{+}$ & 328.7&E2&100.0(16)&1.59(3)&+0.33(3)&-&-&-\\

4$^{+}$ & 2$^{+}$ & 507.5&E2&64.2(11)&1.57(4)&+0.31(8)&-&-&-\\

6$^{+}$ & 4$^{+}$ & 627.8&E2&61.1(11)&1.59(4)&+0.02(1)&-&-&-\\

8$^{+}$ & 6$^{+}$ & 675.9&E2&60.1(11)&1.63(4)&+0.16(2) & 1.2(2)  &0.5(1)  &3.9(3) \\

10$^{+}$ & 8$^{+}$ & 657.2&E2&35.5(7)&1.59(5)&+0.03(2) & 0.9(1)  &0.7(1)  &4.6(3) \\

12$^{+}$ & 10$^{+}$ & 470.9&E2&18.7(4)&1.63(7)&+0.06(3)& 3.5(4)  &0.8(1)$\footnotemark[4]$  &4.8(3) \\

14$^{+}$ & 12$^{+}$ & 582.0&E2&17.1(3)&1.58(8)&+0.22(8)& 1.5(1)  &0.8(1)  &4.8(3) \\

16$^{+}$ & 14$^{+}$ & 743.4&E2&8.1(5)&1.63(8)&+0.36(4) & 0.6(1)  &0.6(1)  &4.1(3) \\

18$^{+}$ & 16$^{+}$ & 868.1&E2&5.9(4)&1.59(16)&+0.17(6)& 0.6$\downarrow$  &0.3$\uparrow$  &2.9$\uparrow$ \\

(20$^{+}$) & 18$^{+}$ & 961.5&(E2)&4.0(3)&-&-&-&-&-\\



\end{tabular}
\end{ruledtabular}
\footnotetext[1]{Uncertainty in $ \gamma $-ray energy is $\pm$ (0.3-0.5) keV.}
\footnotetext[2]{Intensities are normalized with the intensity of the 328.7-keV (2$^{+}$ $\rightarrow$ 0$^{+}$) $\gamma$-transition as 100.}
\footnotetext[3]{Uncertainty due to stopping powers which may be as large as 10\% of the measured lifetimes has not been included in the quoted errors.}
\footnotetext[4]{$B(E2)$ value is calculated by considering the 82(4)\% branching ratio of the 470.9 keV transition \cite{nnd3} measured from the present investigation.}
\end{table*}

The multipolarities of the $\gamma$-ray transitions have been determined from the ADO (Angular Distribution from Oriented nucleus) ratios as described in refs.~\cite{rajban}. In order to obtain the $R_{ADO}$ ratio, two angle-dependent matrices were constructed which stored the coincidence information between the detectors at -23$^\circ$ and 90$^\circ$ and the rest of the detectors. The validity of the method for the present case has been checked from the $R_{ADO}$ values of the known transitions of $^{142}$Gd~\cite{142gd} which was also produced in the present experiment. The ratio would be $0.54(5)$ for pure dipole (823 keV, 5$^{-}$ $\rightarrow$ 4$^{+}$, $E1$) and $1.59(3)$ for quadrupole (793 keV, 6$^{+}$ $\rightarrow$ 4$^{+}$, $E2$) transitions.
Definite parities of the excited states have been designated from the linear polarization asymmetry ratio ($ \Delta_{\text{asym}} $), extracted from the parallel and perpendicular scattering of the $\gamma$ photons inside the detector medium, as described in refs.~\cite{114te,197tl,198bi}. The correction due to the asymmetry in the array and the response of the Clover segments, defined by $a(E_\gamma$) = $\frac{N_\parallel} {N_\perp}$, was estimated using the $^{152}$Eu source and was found to be 0.96(1). The validity of this method has been confirmed from the known transitions of $^{142}$Gd~\cite{142gd} which was also produced in the same experiment. 

A partial level scheme of $^{140}$Gd, as obtained in the present work, is shown in Fig.~\ref{LS}. 
The experimental results obtained in the present work were summarized in Table~\ref{lifetime}.


Lifetimes of the states have been extracted using the LINESHAPE package \cite{wells, johnson}, by analyzing the experimental Doppler-broadened shapes of the $\gamma$-ray transitions in the 
band structure QB I (ground state band) in $^{140}$Gd. Shell-corrected tabulations of Northcliffe and Schilling were used to calculate the energy loss of the residual nuclei when interacting with the atomic electrons of the target-backing combination. These simulated trajectories were then used to generate velocity profiles of the residual nuclei ($^{140}$Gd) at different angles by assuming the response of a composite clover detector was identical to a high-purity single germanium-crystal (HPGe) detector with the dimension same as the former placed at the same position \cite{rajban1}. The velocity profiles were used as the input parameter to calculate the expected Doppler shapes of the $\gamma$ transitions. The lifetimes were finally determined by the least square fitting of the experimental (gated) spectra with the calculated shapes. In the fitting procedure of extracting the lifetime from the experimental spectrum created from a gate on a transition below the $\gamma$ ray of interest, it was necessary to incorporate the contributions from the side feeding transitions. In the present analysis, the side feeding was modelled with a cascade of five transitions and the moment of inertia of the sequence was typically chosen to be similar to that of the bands of interest $i. e$ QB I in $^{140}$Gd. In this process, the effect of variation in the side-feeding intensity between two extreme values (taking the corresponding uncertainties in intensities into account) resulted in a change in the evaluated level lifetime by less than 10\%. The detailed analysis procedure was discussed in Refs. \cite{rajban,rajban1}.

\begin{figure}[t]
\centering
\setlength{\unitlength}{0.05\textwidth}
\begin{picture}(10,7.2)
\put(-0.4,-0.50){\includegraphics[width=0.54\textwidth, angle = 0]{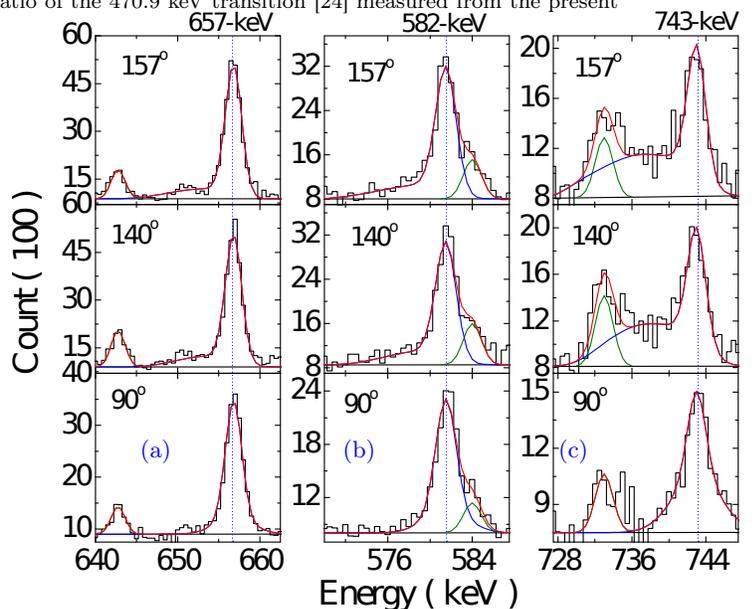}}

\put(1.5,1.8){\textcolor{blue}{(a)}}
\put(4.5,1.8){\textcolor{blue}{(b)}}
\put(7.7,1.8){\textcolor{blue}{(c)}}

\end{picture}
\caption{\label{linsh} The experimental spectra along with the fitted line shapes for the $\gamma$-ray transitions of energy (a) 657 and (b) 582 and (c) 743-keV of QB I in $^{140}$Gd. The obtained line-shape, contaminant peaks, and total line-shapes of the $\gamma$ transitions are represented by the blue, green, and red lines, respectively.}
\end{figure}

The 868.1-keV (18$^{+}$ $\rightarrow$ 16$^{+}$) transition is the topmost $\gamma$-ray transition of the QB I for which a clear shape was observed, wherefrom the effective lifetime of 0.61 ps has been extracted for the 18$^{+}$ state. This information was then used as the input parameter for extracting the lifetimes of the lower-lying levels of the band QB I. Table \ref{lifetime} summarizes the lifetime results along with the $B(E2)$ and $Q_{t}$ values of the QB I in $^{140}$Gd. The observed Doppler broadened shape in the experimental spectra has been fitted simultaneously at three different angles 90$^{\circ}$, 140$^{\circ}$ and 157$^{\circ}$ to obtain the lifetime of the states results as illustrated in Fig. \ref{linsh}. The uncertainties on the quoted lifetimes were derived from the behaviour of the chi-squared value in the vicinity of the least square minimum. The uncertainties do not include the systematic contribution of the stopping powers that has been expected to be $\sim$ 10\%.

\section{Discussions}

The nuclei around $A$ $\approx$ 140 in the light rare-earth region are of interest since they lie on the edge of a predicted region of well deformed prolate rotors. The low-lying yrast states of nuclei in this region have previously been investigated where-from lifetime measurements confirm that the low-lying states show a trend to a well deformed nuclear shape with of the order of $\varepsilon_2$ $\sim$ 0.2 for nuclei with $N$ $\sim$ 78 and $Z$ $\sim$ 64.

\subsection{CNS}

{The origin and characteristics of the nuclear structure have been probed 
through the framework of cranked
Nilsson-Strutinsky (CNS) model. It is to be mentioned that pairing interaction has been neglected in this model \cite{afana,carlsson,bengtsson}. }
In this model, the configurations are designated by
the number of particles or holes in orbitals labeled by the $N$ oscillator
shell, which is further categorised into high- and low- $j$ shells. The rotational symmetry of static nuclear potential is broken when exposed to cranking, thereby making the signature quantum number $\alpha$ = 1/2 or -1/2 a good quantum number for any nuclear state.
The calculations are performed with $\kappa$ and $\mu$
parameters fitted for the A = 150 region that defines the contribution of spin-orbit coupling and an anharmonic term in the modified oscillator potential~\cite{afana, nilsson}. The total energy is calculated as a
sum of the rotating liquid drop energy and the shell energy. This is done 
using the Strutinsky shell correction formalism \cite{andersson, strut}. The
Lublin-Strasbourg drop
(LSD) \cite{pomorsky} is used as a reference for static liquid drop. The rigid-body moment of inertia is calculated
with a radius parameter of $r_0$ = 1.16 fm and diffuseness of
a = 0.6 fm \cite{carlsson}. The calculations minimize the total energy
for the different configurations with respect to deformation parameters,
$\varepsilon_2$, $\varepsilon_4$, and $\gamma$ , at different angular momenta. The configurations are labeled
as per the nomenclature : $ [p_1, n_1]$, where, where $p_1$ and $n_1$ are the numbers of protons and
neutrons in $h_{11/2}$ orbital respectively. A subscript + or - is often used with odd occupation
numbers to highlight the signature, $\alpha$ = 1/2 or -1/2, respectively. 

\begin{figure}
\begin{center} 
\includegraphics[trim=2.5cm 0 4.5cm 0, clip=true,width=3.9cm,angle=0]{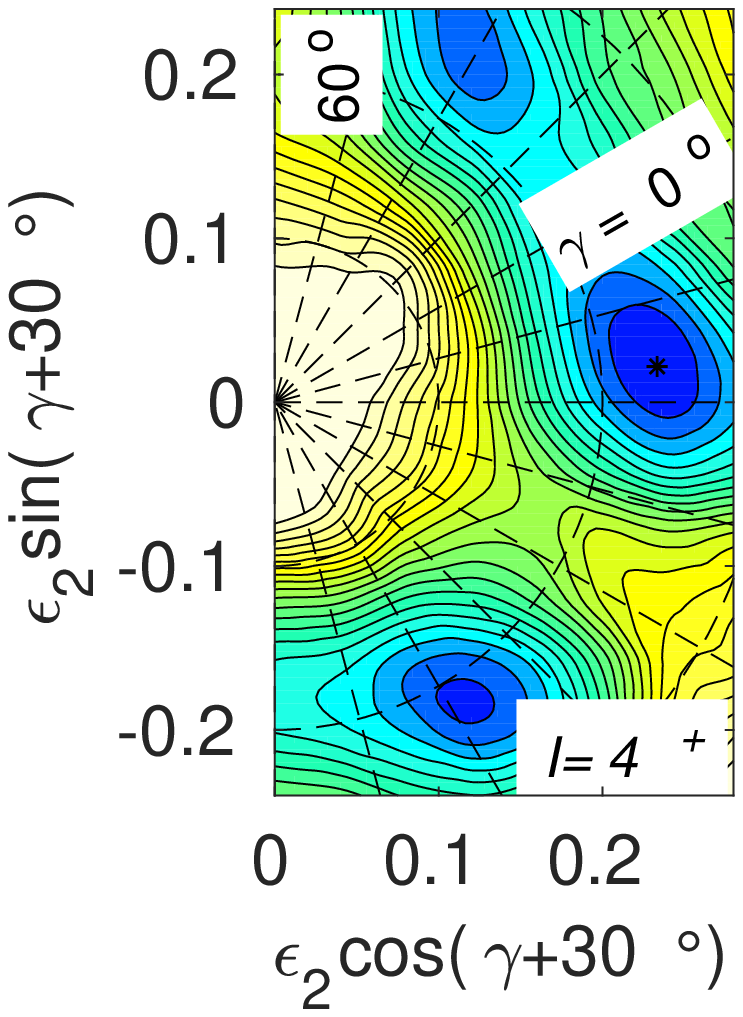}
\includegraphics[trim=2.5cm 0 4.5cm 0, clip=true,width=3.9cm,angle=0]{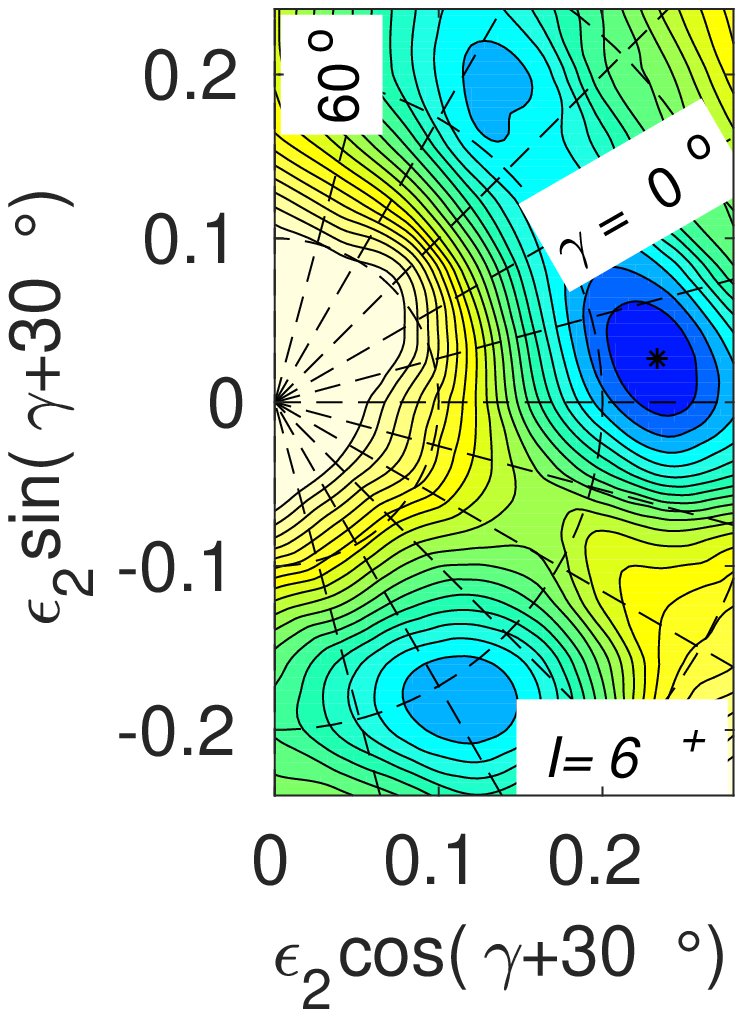}\\
\vspace{0.2cm}
\includegraphics[trim=2.5cm 0 4.5cm 0, clip=true,width=3.9cm,angle=0]{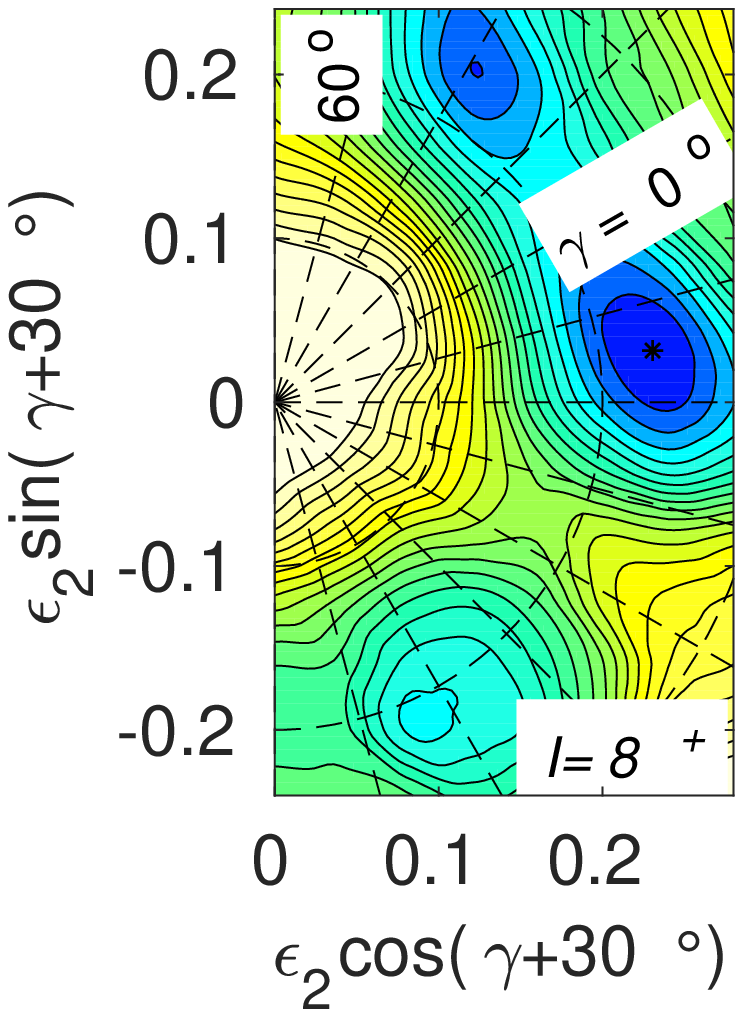}
\includegraphics[trim=2.5cm 0 4.5cm 0, clip=true,width=3.9cm,angle=0]{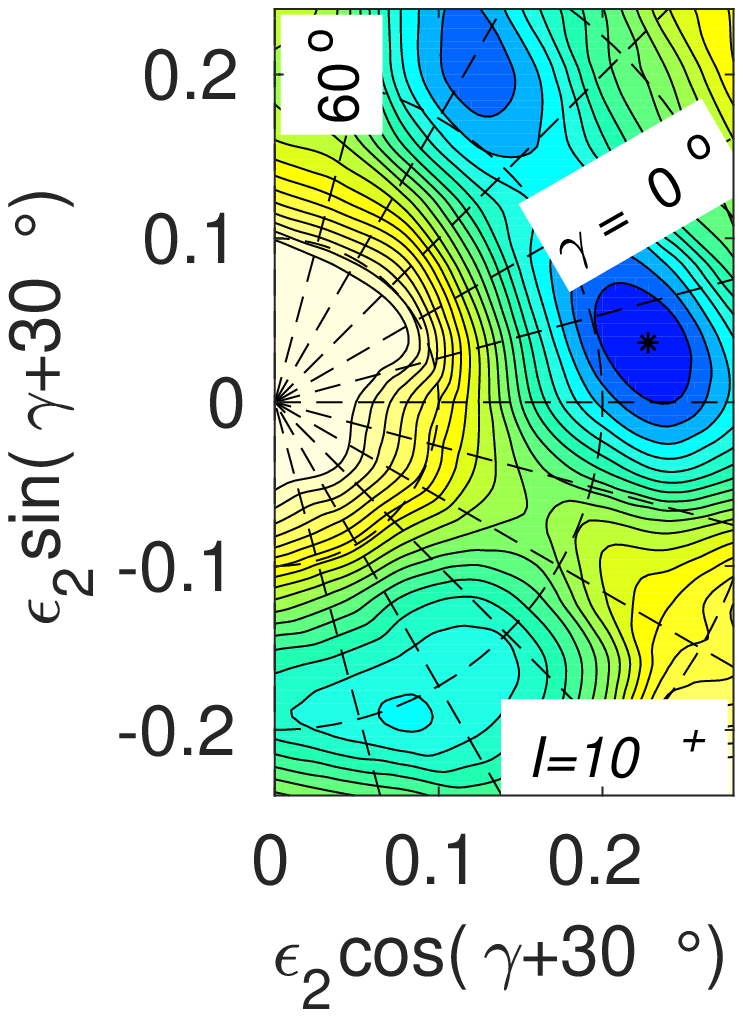}\\
\vspace{0.2cm}
\includegraphics[trim=2.5cm 0 4.5cm 0, clip=true,width=3.9cm,angle=0]{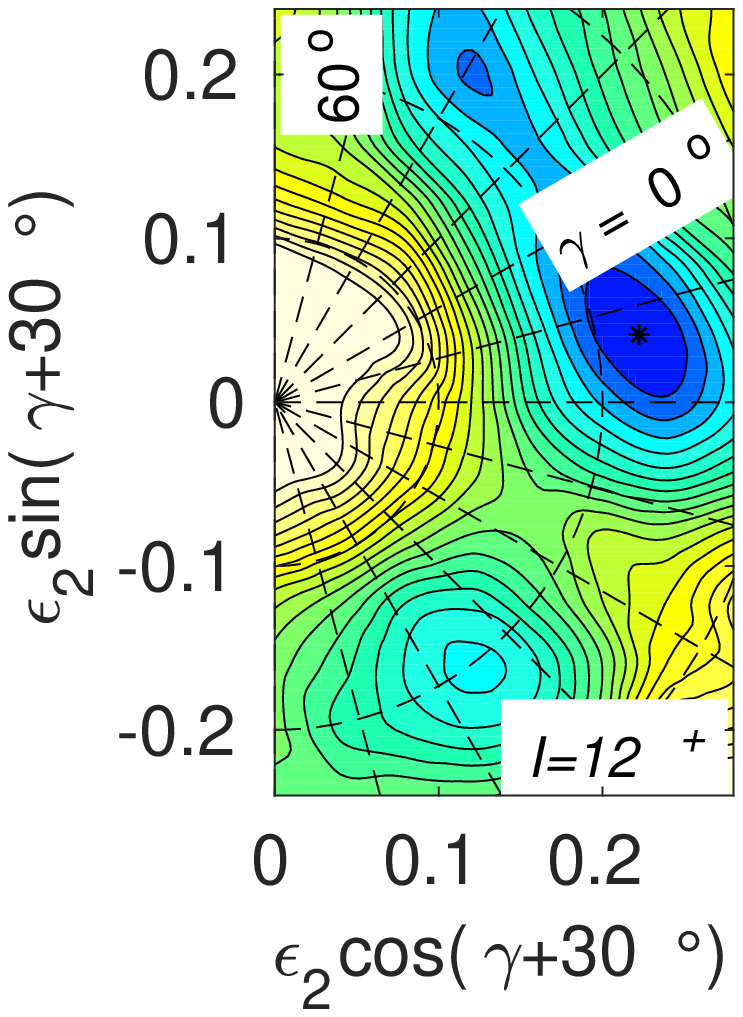}
\includegraphics[trim=2.5cm 0 4.5cm 0, clip=true,width=3.9cm,angle=0]{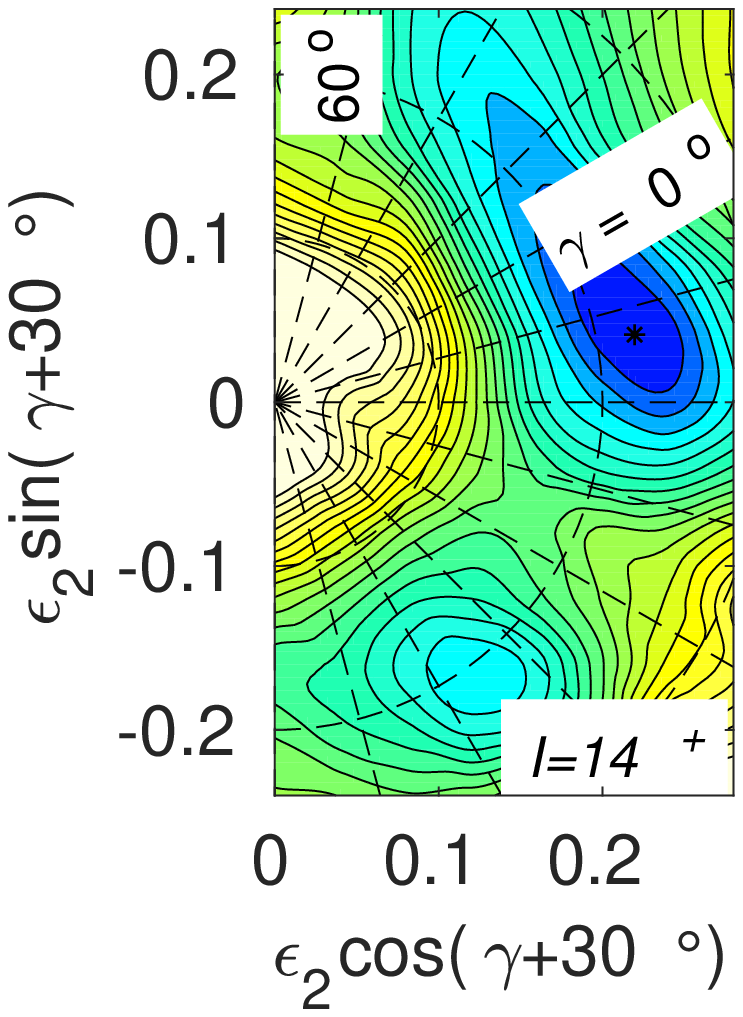}\\     
\vspace{0.2cm} 
\includegraphics[trim=2.5cm 0 4.5cm 0, clip=true,width=3.9cm,angle=0]{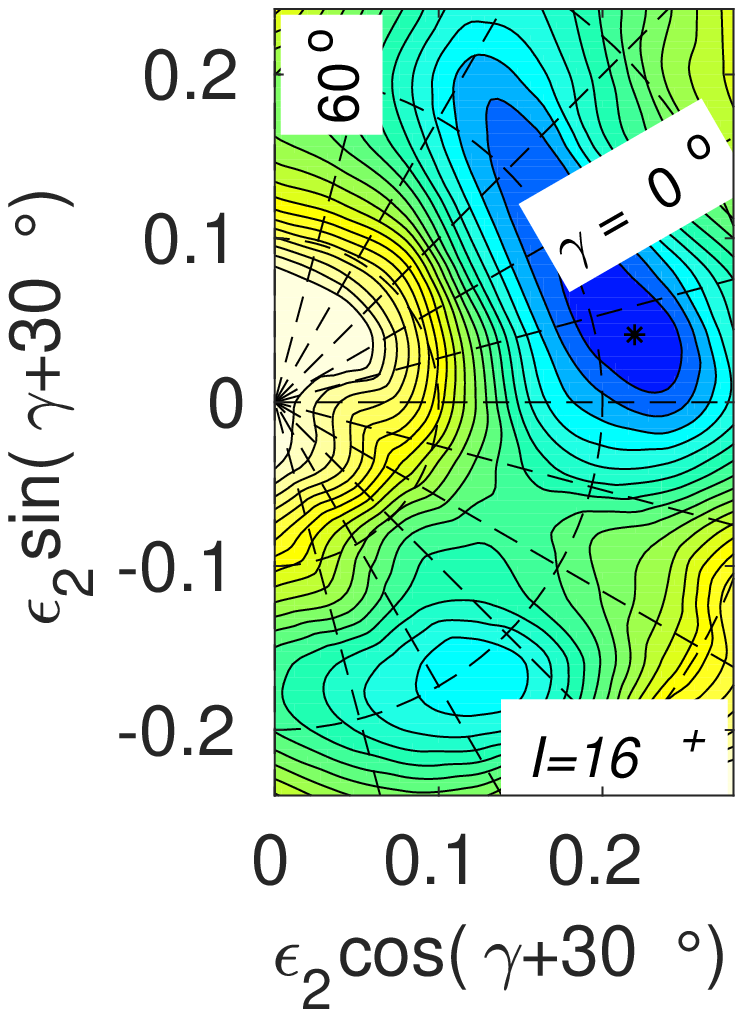} 
\includegraphics[trim=2.5cm 0 4.5cm 0, clip=true,width=3.9cm,angle=0]{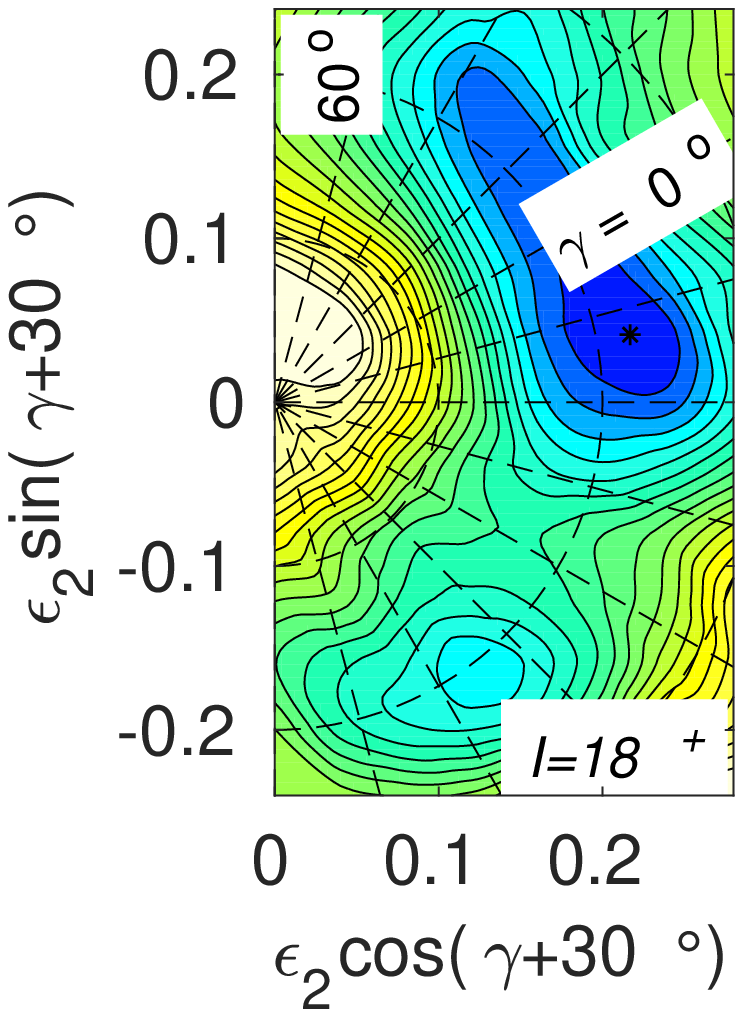}\\ 
\caption{Total energy surfaces for general scan run for spins 4$\hbar$ to 18$\hbar$.
The contour line separation is 0.25 MeV.}
 \label{pes06} 
\end{center}
\end{figure}

\begin{figure}
\begin{center} 
\includegraphics[trim=2.5cm 0 4.5cm 0, clip=true,width=3.9cm,angle=0]{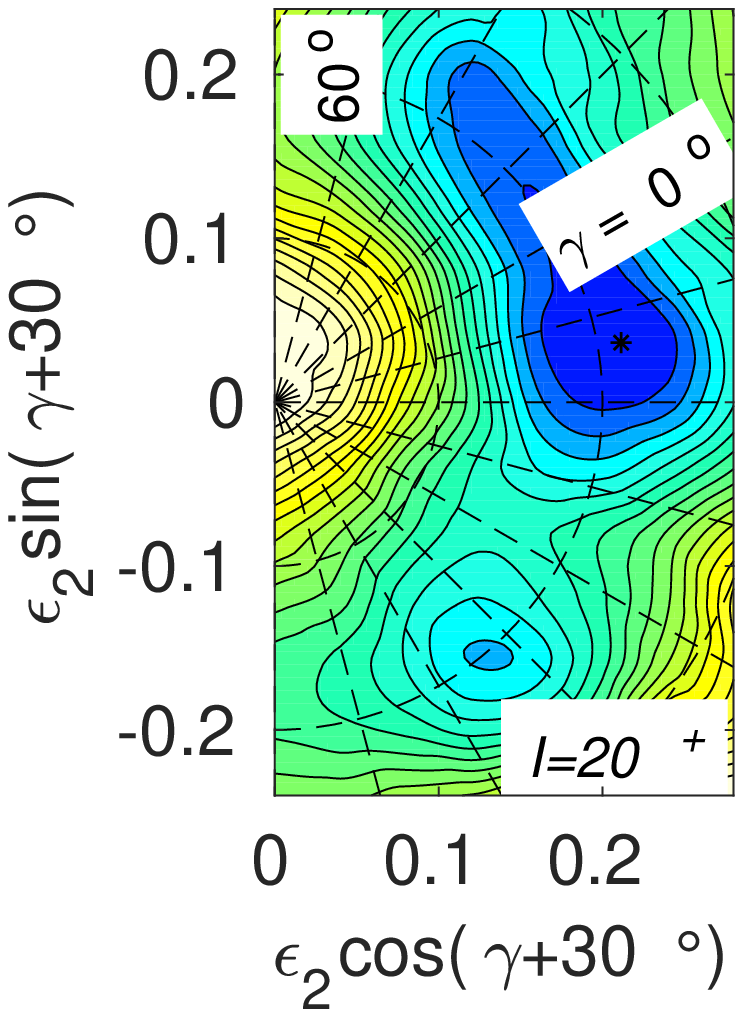} 
\includegraphics[trim=2.5cm 0 4.5cm 0, clip=true,width=3.9cm,angle=0]{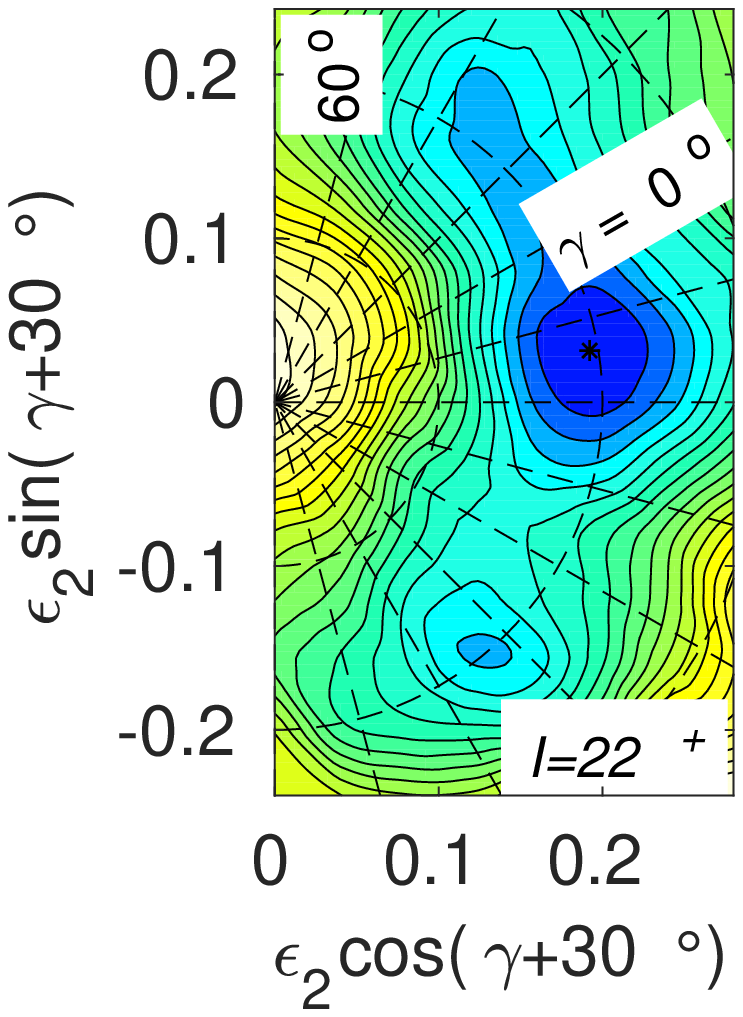} \\
\includegraphics[trim=2.5cm 0 4.5cm 0, clip=true,width=3.9cm,angle=0]{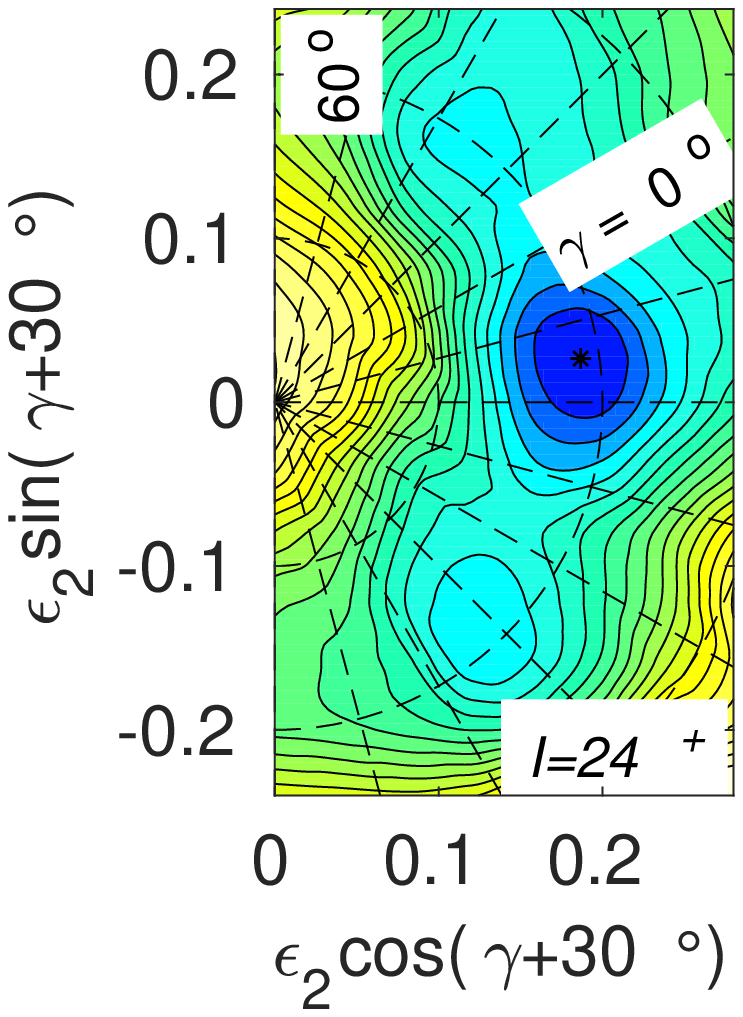} 
\includegraphics[trim=2.5cm 0 4.5cm 0, clip=true,width=3.9cm,angle=0]{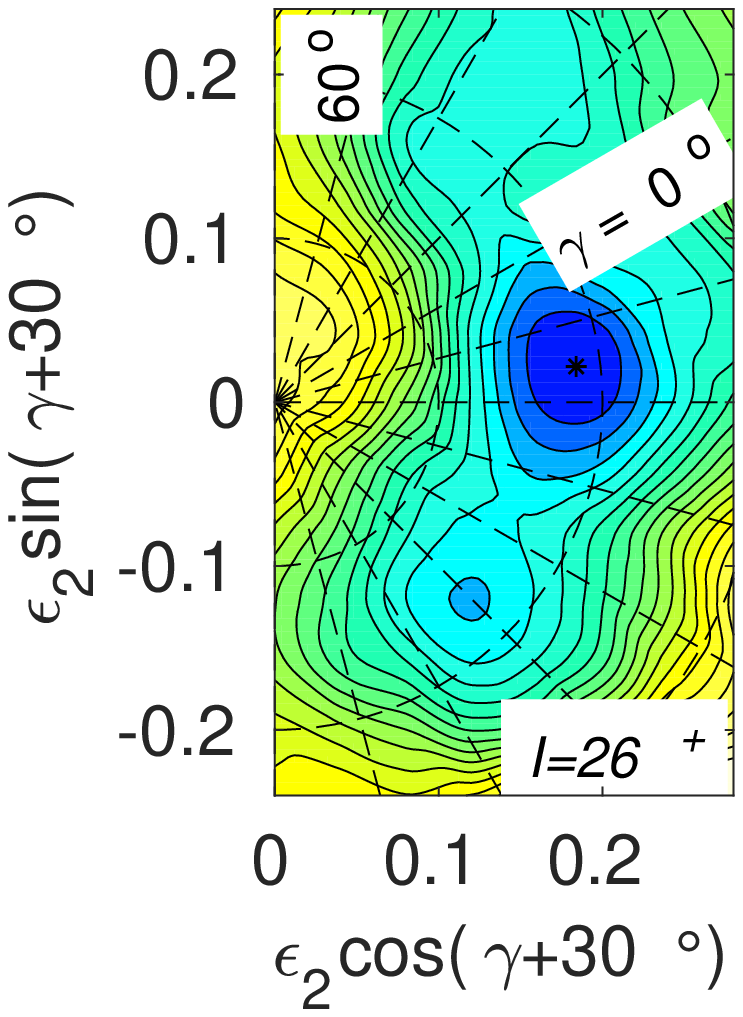} \\
\vspace{0.2cm}
\includegraphics[trim=2.5cm 0 4.5cm 0, clip=true,width=3.9cm,angle=0]{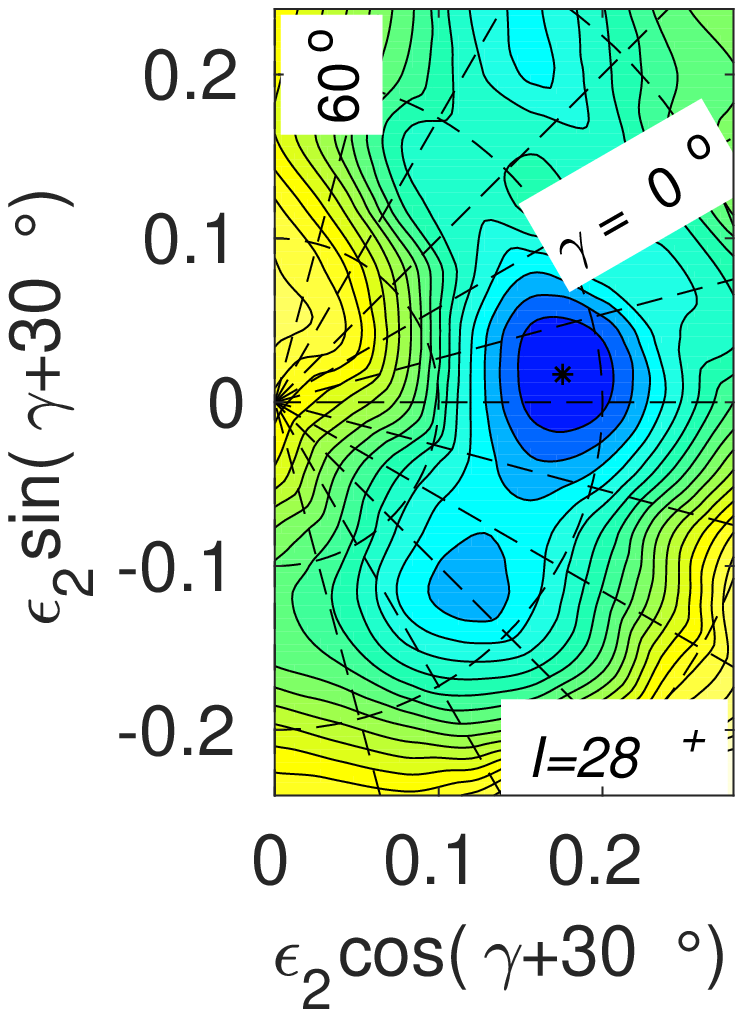} 
\includegraphics[trim=2.5cm 0 4.5cm 0, clip=true,width=3.9cm,angle=0]{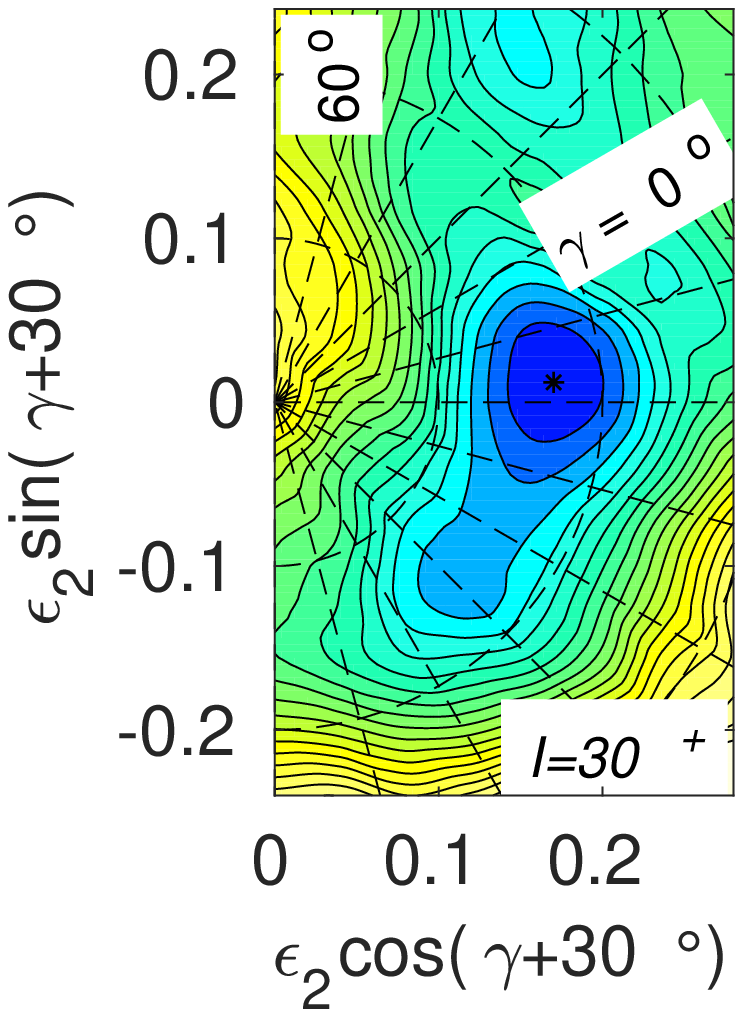}\\
\caption{Total energy surfaces for general scan run for spins 20$\hbar$ to 30$\hbar$.
The contour line separation is 0.25 MeV.}
 \label{pes061} 
\end{center}
\end{figure}



\begin{figure}[ht]
\begin{center}
\includegraphics[scale=0.360,angle=-90]{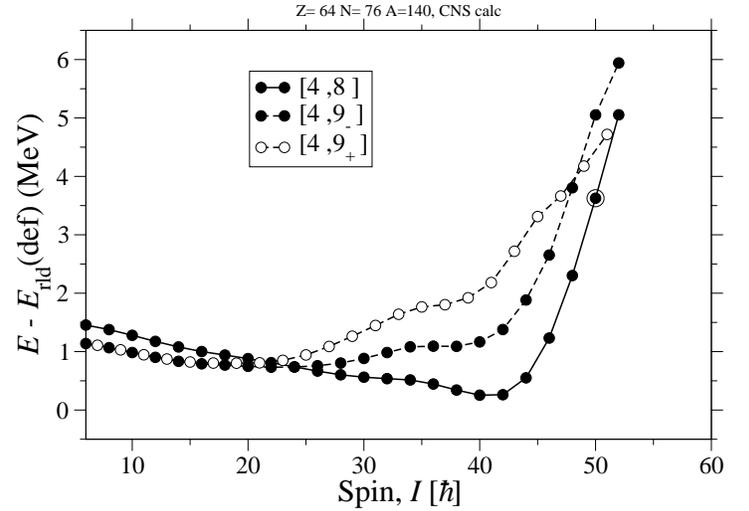}
\caption{\label{erld_fc} {Calculated energies are relative to rotating liquid drop energy as a function of spin for some of the energetically favorable configurations. The negative parity states are shown with dashed lines whereas the positive parity configurations are with continuous lines.}}
\end{center}
\end{figure}

\begin{figure}[ht]
\begin{center}
\includegraphics[scale=0.360,angle=-90]{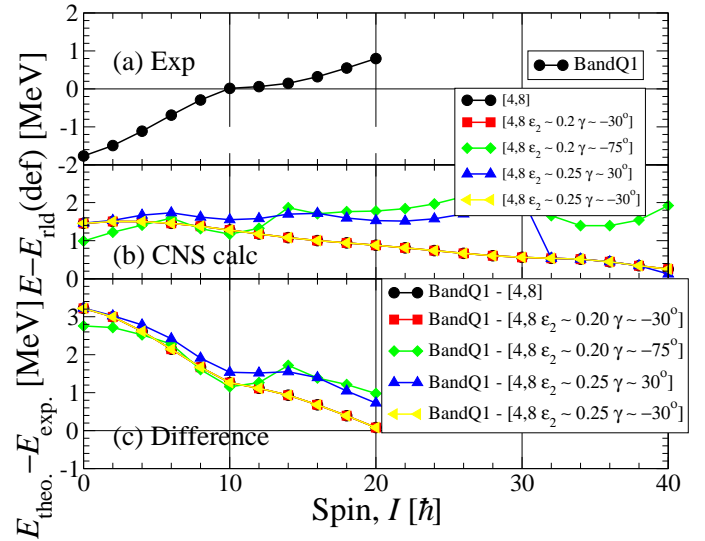}
\caption{\label{erld} (panel a) Excitation energies are relative to a rotating liquid drop 
w.r.t. spin for the observed quadrupole band QB1. (panel b) Calculated energies relative to a rotating liquid drop energy as a function of spin for the configuration {[4,8] at different specific deformations.}
(panel c) The difference between experiment and theory.}
\end{center}
\end{figure}

\begin{figure}
\begin{center} 
\includegraphics[clip=true,width=3.0cm,angle=-90]{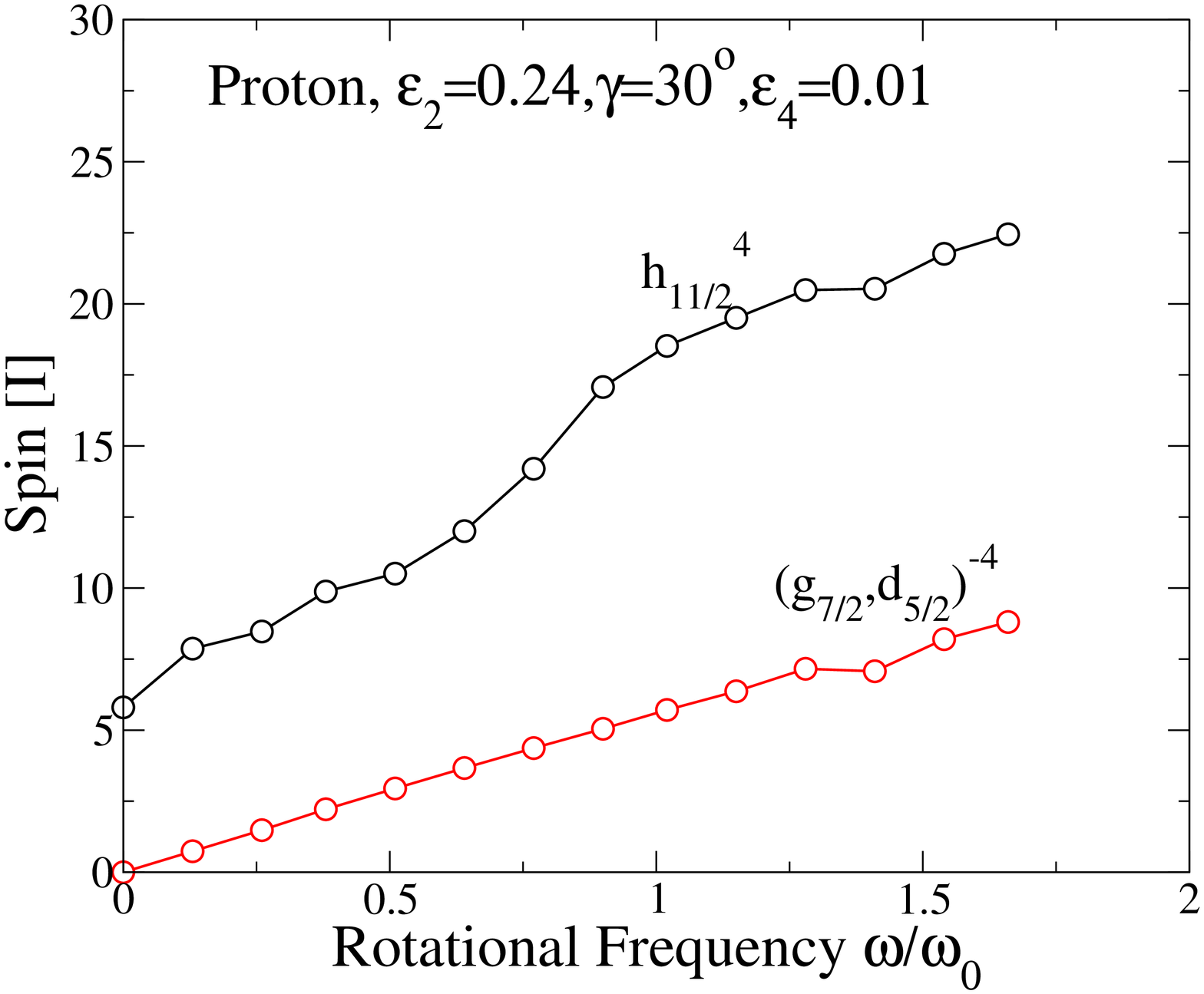}
\includegraphics[clip=true,width=3.0cm,angle=-90]{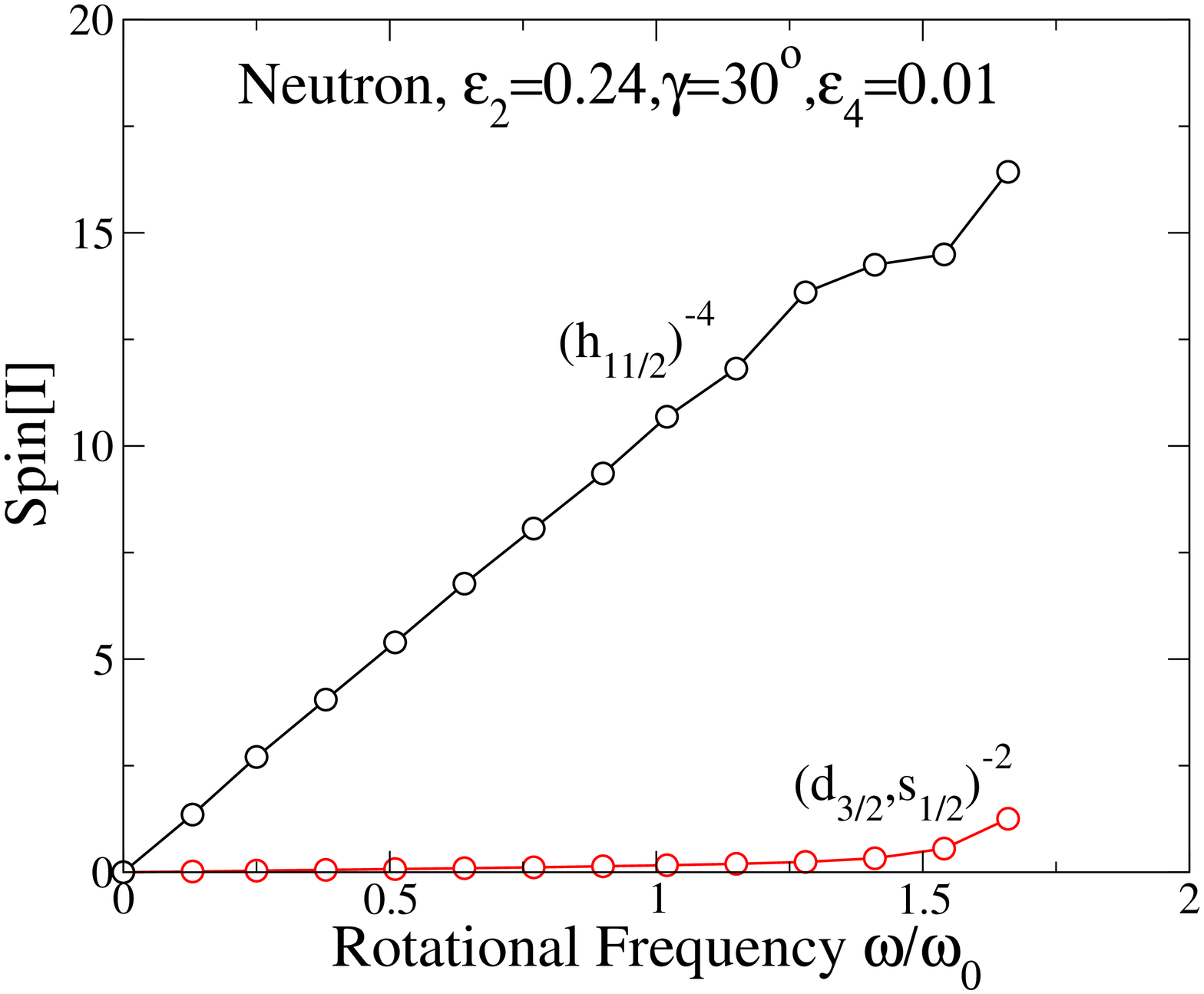}\\
\vspace{0.1cm}
\includegraphics[clip=true,width=3.0cm,angle=-90]{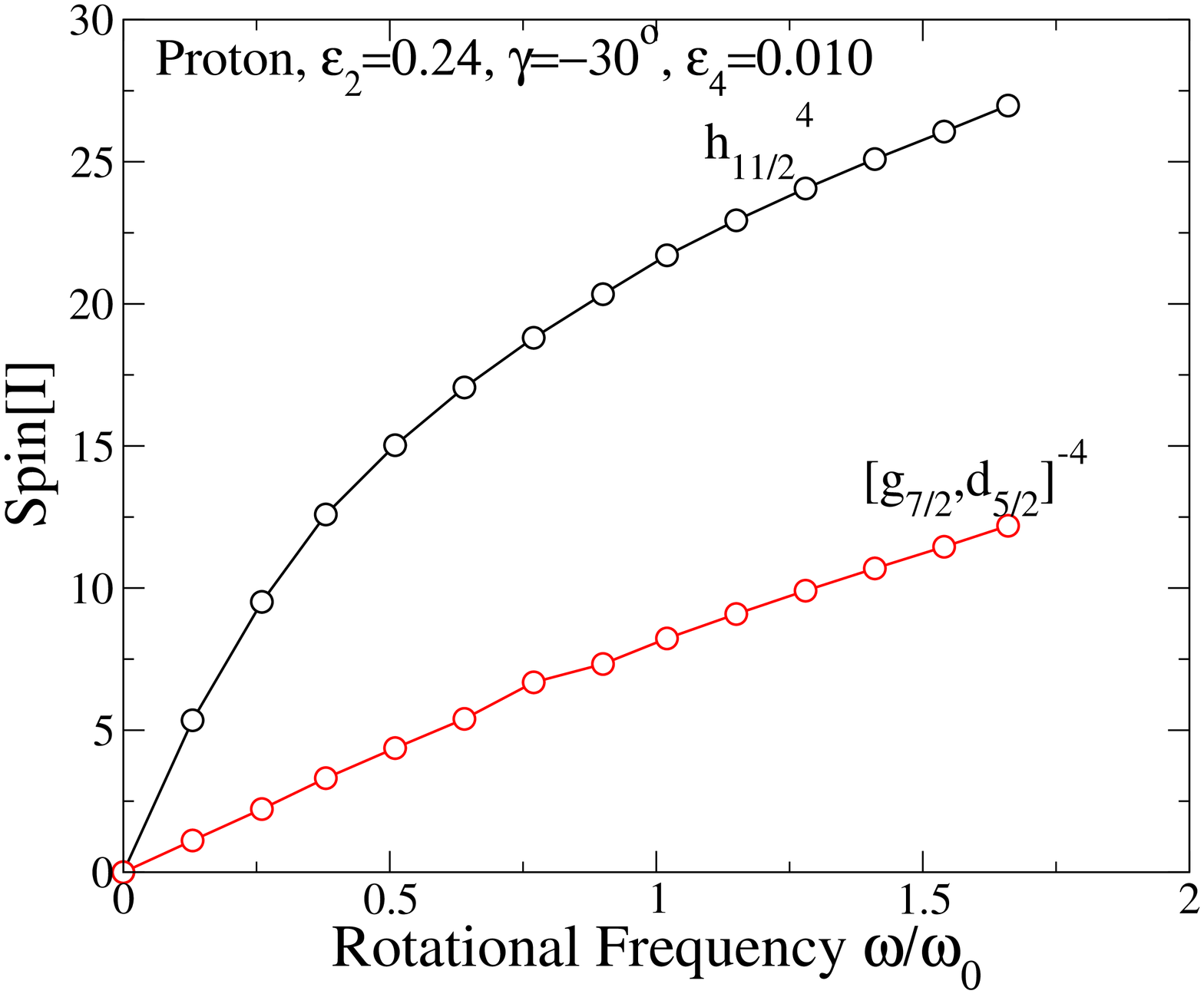}
\includegraphics[clip=true,width=3.0cm,angle=-90]{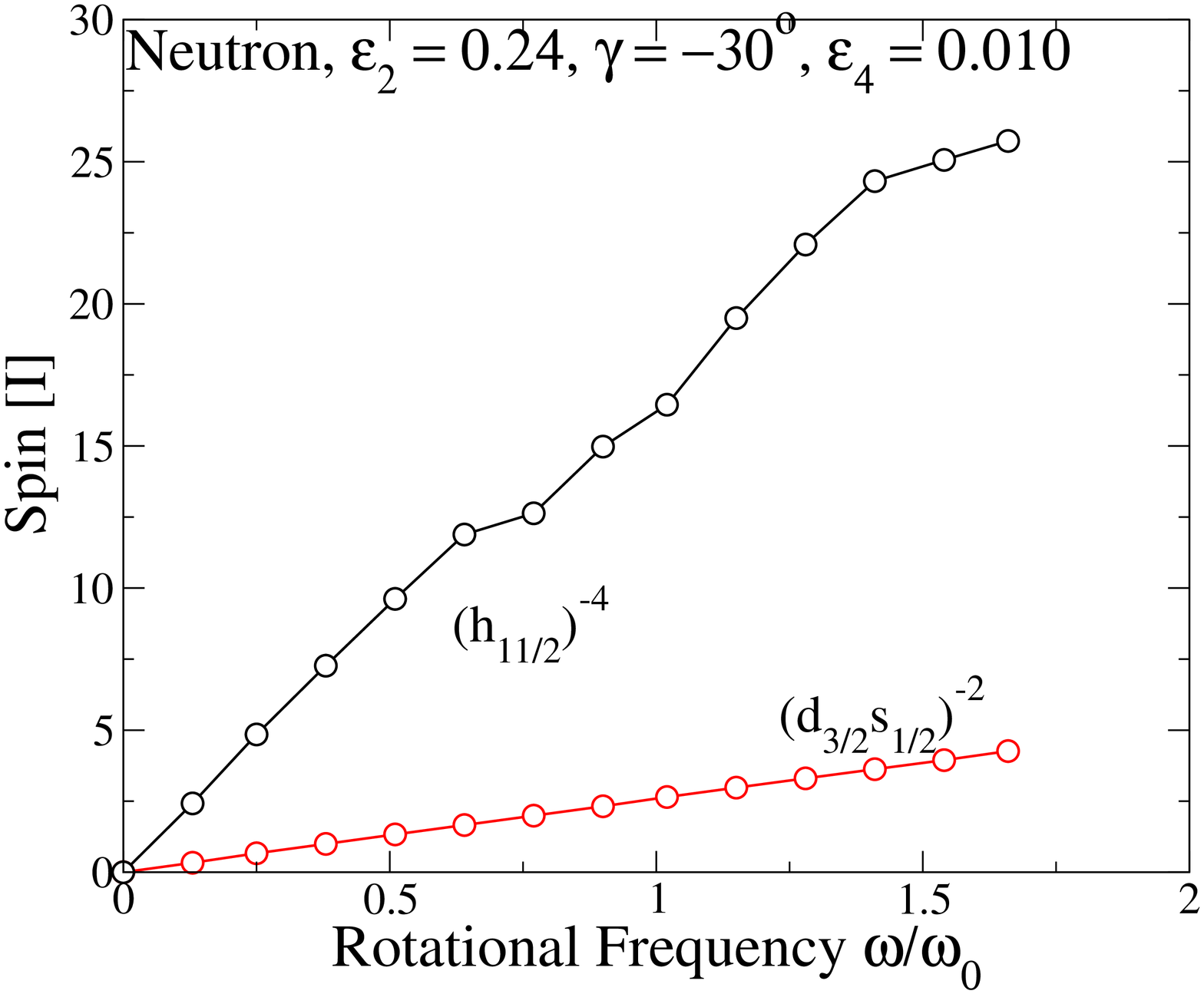}\\
\vspace{0.1cm}
\includegraphics[clip=true,width=3.0cm,angle=-90]{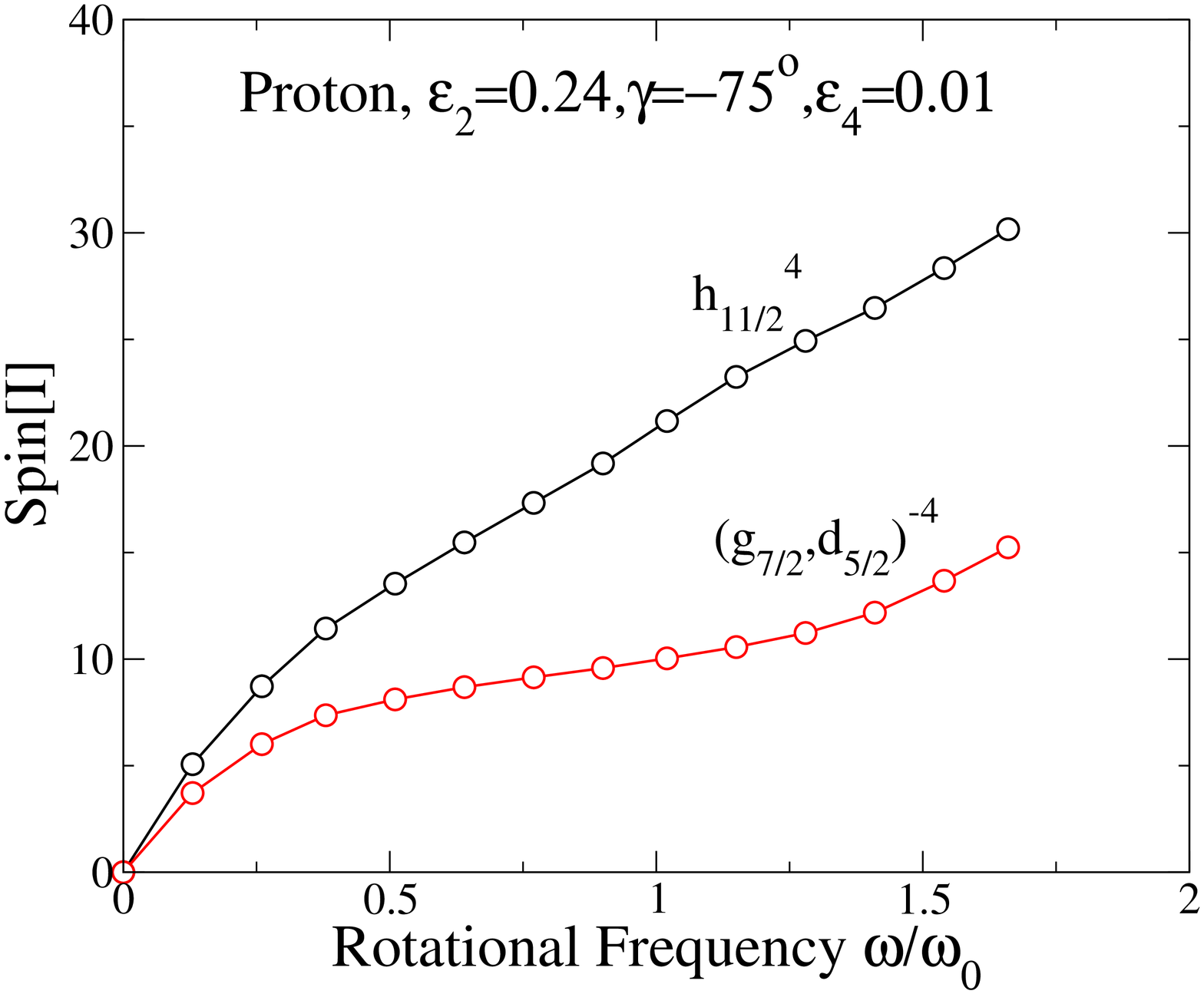}
\includegraphics[clip=true,width=3.0cm,angle=-90]{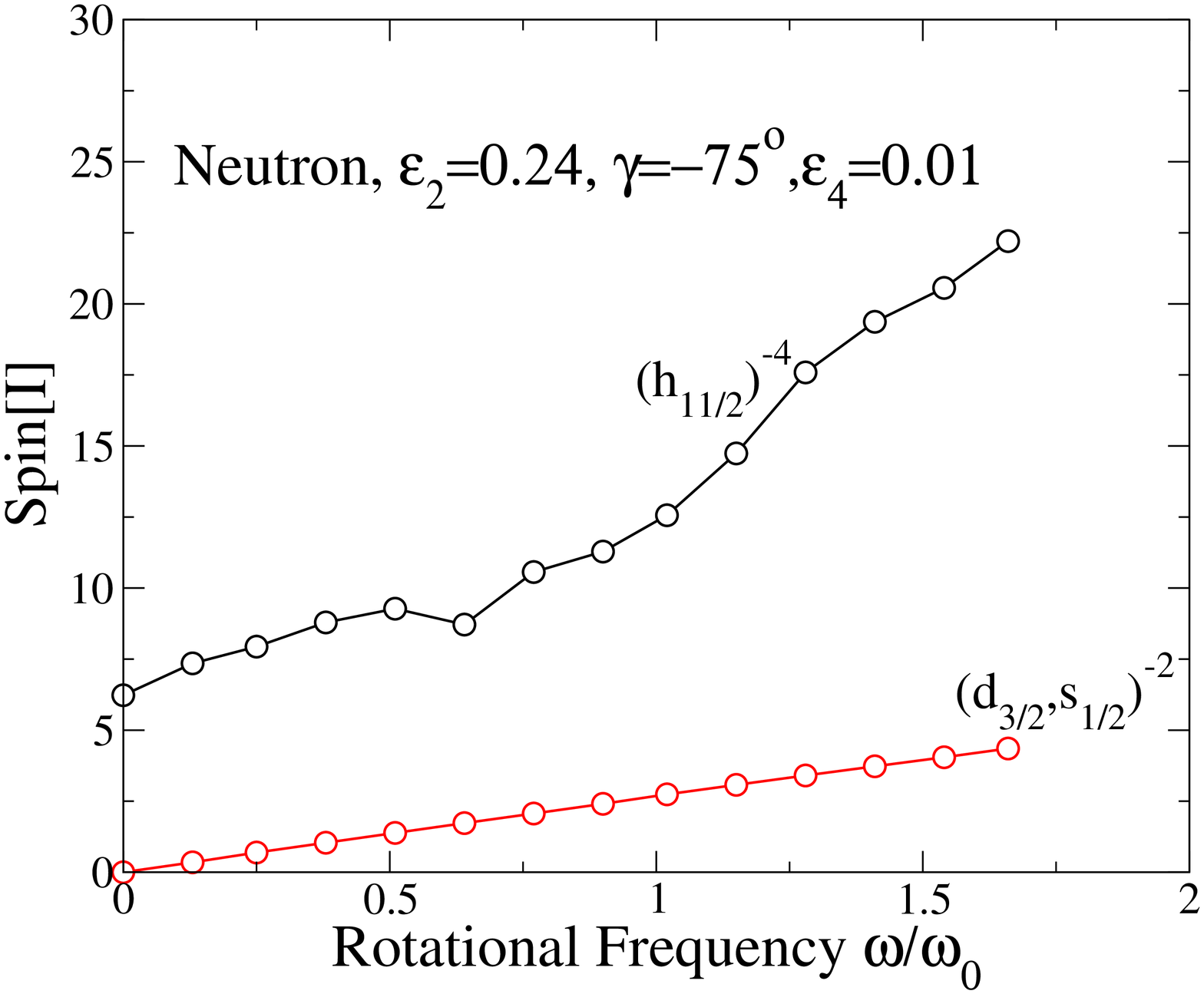}\\
\newpage
\caption{(Color online) Alignment for the different groups of
orbitals of the (unpaired) [4,8] configuration at a typical deformation
($\varepsilon_2$ = 0.24) and for rotation around the longest ($\gamma$ =
$-$75$^\circ$, the intermediate ($\gamma$ = $-$30$^\circ$ ) and the shortest ($\gamma$ = 30$^\circ$ )
principal axis. The frequency is expressed in oscillator units and
should be multiplied by $\sim$ 10 to get $\hbar\omega$ in MeV. The alignment is
shown for the four $g_{7/2}d_{5/2}$ proton holes and two $d_{3/2},s_{1/2}$ neutron holes
by red color symbol and the total alignment of the valence protons
and neutrons are shown by the black color symbol. Consequently, the alignment
of the 4 $h_{11/2}$ protons and the 4 $h_{11/2}$ neutron holes, respectively,
is the difference between the dashed and full lines.}
 \label{align} 
\end{center}
\end{figure}

The evolution of the different minima (Fig. \ref{pes06} and \ref{pes061}) with spin is shown in the given ($\varepsilon , \gamma$) mesh. This is subject to how the angular
momenta of particles and holes get aligned along with the rotation of the nucleus. Thus, when particles align along one of the principal {axes},
the matter distribution gets concentrated around the
equator. This leads to a rotational phenomenon around the
shortest principal axis. On the other hand,
the aligning of holes will result in the removal of matter from the equatorial plane and thus
lead to the rotation of the nucleus around the longest
principal axis. At spin $I \sim$ 4$\hbar$, we observe three minima with $\gamma$ $\sim$ +30$^\circ$, $-$30$^\circ$ and $-$75$^\circ$ respectively. 
The minima at $\gamma$ $\sim$ 30$^\circ$ correspond to rotation around the shortest axis whereas that at $\gamma$ $\sim$ $-$30$^\circ$ represents rotation around the intermediate axis. The rotation around the longest principal axis would be generated with $\gamma$ $\sim$ $-$75$^\circ$. For details one can refer to the \cite{pmoller,cmpetra,rbengts,tseinha,bgcarls,cmpetr222}. We will now trace these minima and would try to understand the evolution of nuclear shapes with an increase in angular momenta. The rotation around the longest axis ceases to exist by the time $I \sim$ 10$\hbar$ is reached. Thereafter rotations around the intermediate and the shortest principal {axes} become dominant till $I \sim$ 20$\hbar$. Beyond this, the minimum at $\gamma$ $\sim$ +30$^\circ$ gradually traverses towards $\gamma$ $\sim$ $-$30$^\circ$. As a result rotation around the intermediate axis prevails alone till $I \sim$ 26$\hbar$ is reached  where re-evolution of $\gamma$ $\sim$ $-$75$^\circ$ is observed. It is to be mentioned that the latter $\gamma$ $\sim$ $-$75$^\circ$ is produced at comparatively lower axial deformation of $\varepsilon \sim$ 0.2 compared to the earlier case. So, now it remains to be calculated which configurations would correspond to the above shapes at the given spin values. As explained above, several configurations were calculated through minimization of energy over a mesh defined by 
$\varepsilon_2$, $\varepsilon_4$ and $\gamma$ parameters. {Among those, some of the energetically favorable configurations are selected and compared to the observed bands in $^{140}$Gd (see Fig. \ref{erld_fc}). We found that the configuration {[4, 8]} comes out to be the yrast. Excitation energy with respect to liquid drop for experimental data is shown in Fig. \ref{erld} (panel a).  We probed in to the configuration at around fixed deformation of $\gamma$ $\sim$ {$-$75$^\circ$}, {$-$30$^\circ$} and 30$^\circ$ (see Fig. \ref{erld} (panel b) and compared the excitation energy with respect to rotating liquid drop versus spin in the configuration {[4, 8]} at different triaxial deformation parameter (see Fig. \ref{erld}(panel c)).} The present {[4,8]} configuration of $^{140}$Gd has
more holes than particles so that the tendencies for rotation
around the longest axis becomes preferable. Thus, rotation with the smallest $J_{rig}$,
becomes dominating for this type of valence configuration.


The configurations in Fig. \ref{align} have been chosen at an inter-
mediate frequency, $\omega/\omega_0$ = 0.038 and are tracked continuously to small and high frequencies. It is to be noted that the $\varepsilon_4$ has been fixed at 0.01. The explicit calculations for the {$\pi (h_{11/2})^4 \otimes \nu (h_{11/2})^{−4}$} configuration with protons at $\gamma$ = {$-$75$^\circ$}, the four $g_{7/2},d_{5/2}$
proton holes come close to their full alignment (10 $\hbar$) at a
low frequency, $\omega/\omega_0$ = 0.02, whereas the $h_{11/2}$ particles tend to align gradually to reach 
their full alignment
at the highest frequency. The situation at $\gamma$ = 30$^\circ$ is
just the reverse of this situation, i.e., the four $g_{7/2},d_{5/2}$ holes align gradually,
whereas the four $h_{11/2}$ particles reach full alignment at
a low frequency. On the other hand, for the neutrons at $\gamma$ =
$-$75$^\circ$, the $d_{3/2}, s_{1/2}$ holes align gradually over the full frequency range
even coming above their maximum spin indicating a mixture
from the $d_{5/2}$ and $g_{7/2}$ subshells. The four neutron holes in $h_{11/2}$, however,
are almost fully aligned (16 $\hbar$) in the full frequency range. 
Though at $\gamma$ = 30$^\circ$, we see a gradual increase in neutron hole alignment which aligns the maximum as 
frequency increases.
It is observed that the maximum alignment for four particles in $h_{11/2}$ is more than 16$\hbar$. This could be due to the contribution from intruding $h_{9/2}, f_{7/2}$ orbitals (see discussion below). 
It is also seen that the spin contribution
from the $d_{3/2},s_{1/2}$ neutron holes is almost negative at $\gamma$ = 30$^\circ$. The main feature
of the figure is that high- and intermediate-$j$ holes get fully
aligned at a small frequency for $\gamma$ = {$-$75$^\circ$}
while high- and
intermediate-$j$ particles get fully aligned at a small frequency
for $\gamma$ = 30$^\circ$. Then at $\gamma$ = $-$30$^\circ$ , the collectivity is larger and
all orbitals are seen to align gradually. It should be noted that, for $^{140}$Gd, we do not observed stable $\gamma$ = {$-$89$^\circ$}  shape which is also evident from $E-E_{rld}$ vs spin plot for $\gamma$ = {-90$^\circ$} . But the nucleus $^{142}$Gd was observed to show predominant $\gamma$ = {-90$^\circ$} shape.

\begin{figure}
\begin{center}
\includegraphics[width=6.4cm,  angle = -90]{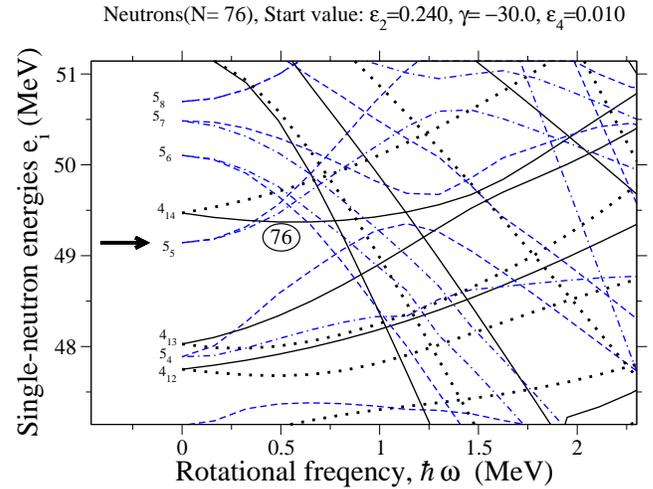} \\ 
\caption{\label{routh} (Color online) Single particle routhians for neutrons (a) at deformation parameters $\varepsilon_2$ = 0.24, $\gamma$ = $-$30$^\circ$ and $\varepsilon_4$ = 0.01. 
Four combinations  of  different ($\pi$,  $\alpha$)  are  plotted  with the 
following representation: solid  lines are for (+, +1/2), dotted  lines represent (+, -1/2)
  whereas  (-, +1/2)  and (-,  -1/2)  are shown by dashed and dashed-dotted lines
  respectively. }
\end{center}
\end{figure}




Now, at $\gamma$ = $-$30$^\circ$ lies in between $\gamma$ = {$-$75$^\circ$} and $\gamma$ = 30$^\circ$. Here, the collectivity is larger and
all orbitals are seen to align gradually.
With the help of routhian plots, let us elaborate onto the
neutron orbitals at $\gamma$ = $-$30$^\circ$ , which are shown in Fig. \ref{routh}. In this
figure, virtual interactions have been removed \cite{virtual}.
Configuration with
N = 76 Fermi level would have four $h_{11/2}$ holes
corresponding to the two strongly upsloping 5$_5$ and 5$_8$ orbitals.
It is to be mentioned that the upslope of the curves (5$_5$ and 5$_8$ orbitals) are proportional to their alignment and because the slope is almost constant over the entire frequency range, the spin contribution is large and almost
independent of rotational frequency as seen in Fig. \ref{align}. At the same time, it is observed that downsloping 5$_6$ and 5$_7$ orbitals intrude into the Fermi level. These correspond to $h_{9/2}, f_{7/2}$ orbitals. The two
N = 4 holes which correspond to $s_{1/2}$ orbitals, however, gradually start to slope upward, leading to a
gradual increase of the spin contribution. There may also be contributions from $d_{3/2}$ orbitals.
If we compare the total spin contribution in the three different cases
( sum of spins of proton and neutron ) it is obvious that
in a large frequency range corresponding to spin $\sim$ 30 $\hbar$ for
$\gamma$ = $-$75$^\circ$, the highest spin at a fixed frequency could only be built
from rotation around the longest principal axis. 

\begin{figure}[ht]
\begin{center}
\includegraphics[scale=0.30,angle=-90]{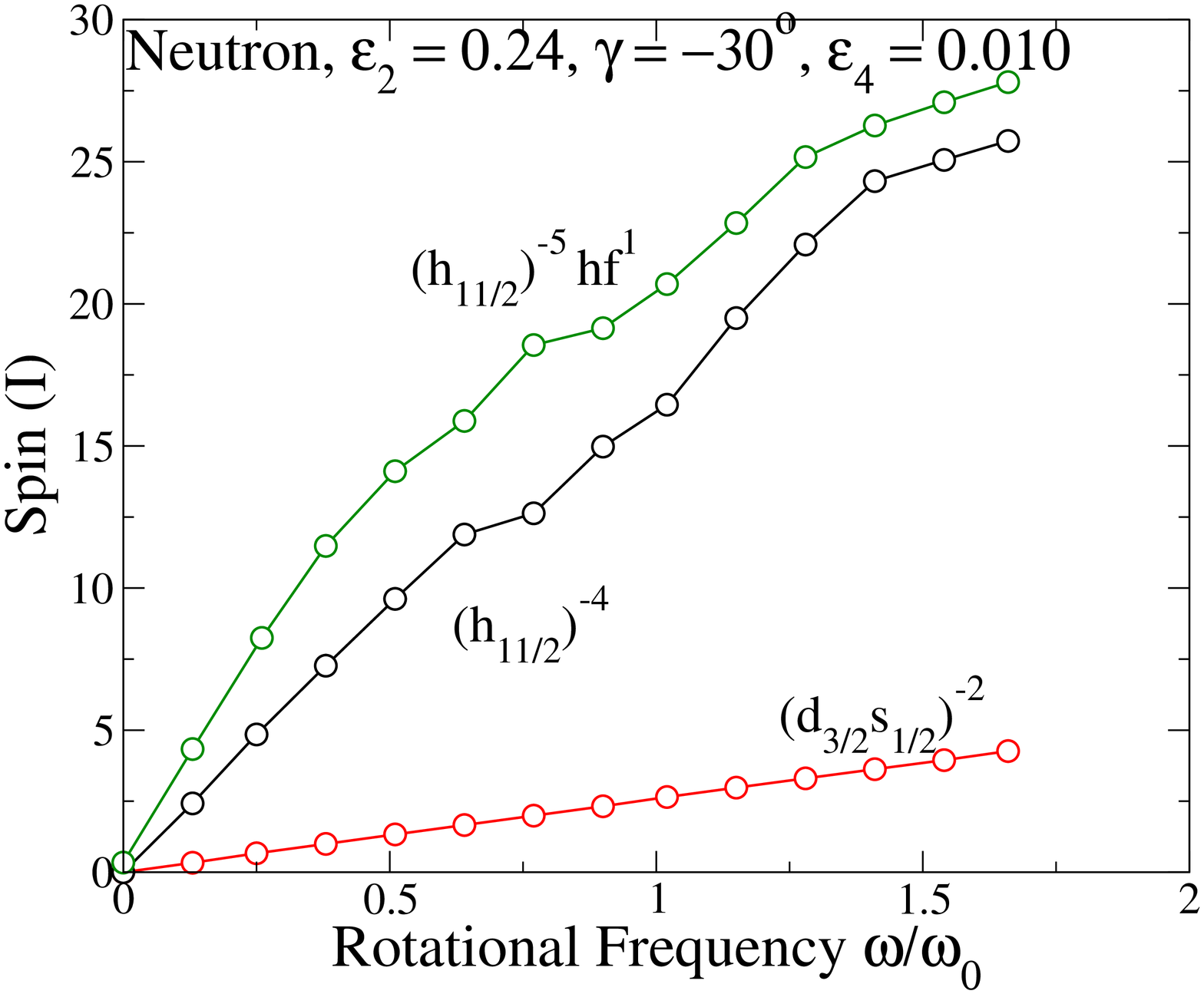}
\includegraphics[scale=0.30,angle=-90]{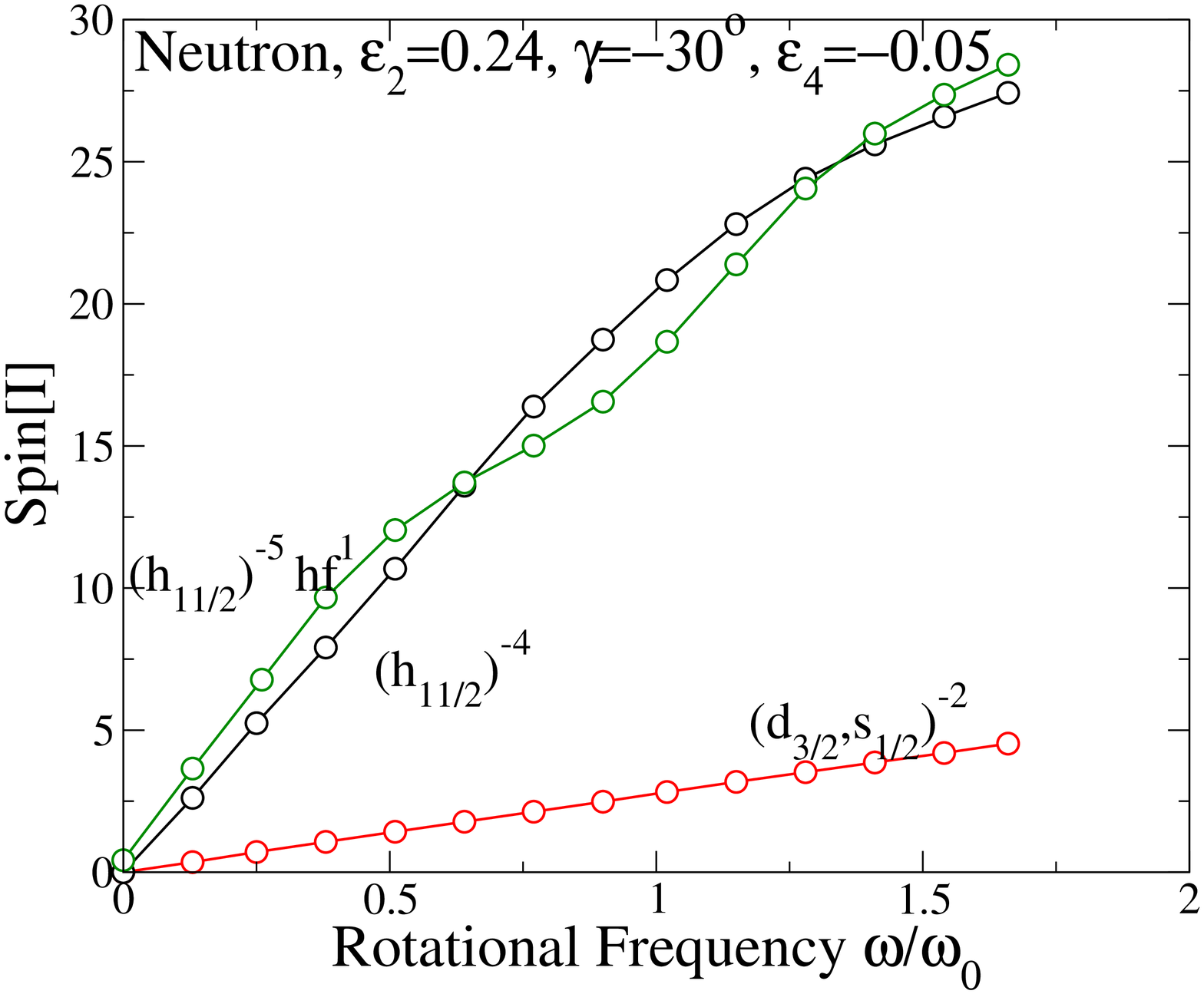}
\caption{\label{alignh9_2} Spin contribution comparison of four neutron holes and five neutron holes coupled with ($hf$) neutrons at (a) $\epsilon_4$ = 0.01 and -0.05 respectively.}
\end{center}
\end{figure}

The intruding orbitals should contribute towards the spin alignments. We compared the spin contributions from $h_{11/2}$ and $h_{9/2}, f_{7/2}$ orbitals for the neutron holes at $\varepsilon_2$ = 0.24, $\gamma$ = $-$30$^\circ$ and $\varepsilon_4$ = 0.01 in Fig. \ref{alignh9_2}. Interestingly, no crossing was observed between configuration with only holes in $h_{11/2}$ and holes in $h_{11/2}$ coupled to $h_{9/2}, f_{7/2}$. On the other hand, we plotted the spin contribution for $\varepsilon_4$ = -0.05 where neutron holes in $h_{11/2}$ coupled to $h_{9/2}, f_{7/2}$ was found to overtake configuration with neutron holes in $h_{11/2}$. So, there is a finite probability that contribution from $(hf)$ neutron orbitals would be predominant beyond frequency = 0.05. 
Therefore, the possible mixing of hole configuration in $h_{11/2}$ and $hf$ cannnot be defined.
At the same time this is not hard and fast constraint on $\varepsilon_4$. 


\begin{figure}[ht]
\begin{center}
\includegraphics[scale=0.35,angle=-90]{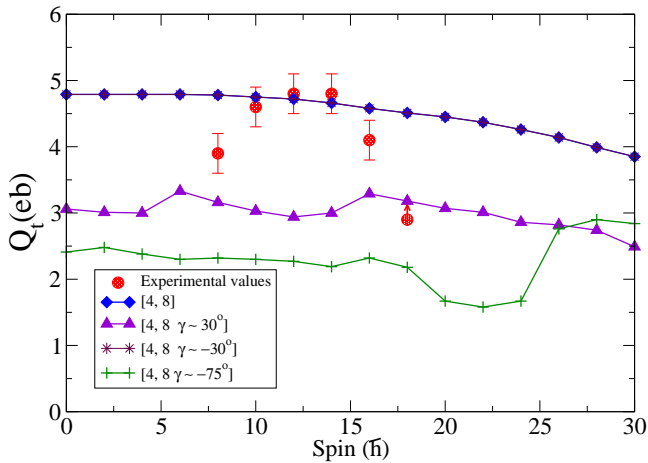}
\caption{\label{qt} Comparison of observed $Q_t$ (shown with red circles along with error bars ) vs spin with those of calculated ones at three given $\gamma$ values.}
\end{center}
\end{figure}

The comparison of observed quadrupole moments as a function of spin with those of calculated ones is shown in Fig. \ref{qt}. The calculated $Q_t$ values come out to be the largest for
the $\gamma \sim$ $-$30$^\circ$ shape, as this is far away from the
non-collective $\gamma$ = $-$120$^\circ$, 60$^\circ$ axis. At lower and intermediate spin
$Q_t$ values calculated at $\gamma \sim$ $-$30$^\circ$ agree well with the
observed values. Thus, it gives a hint of the dominance of rotation of the nucleus around the intermediate 
principal axis. We have only one data point around $I \sim$ 20$\hbar$ which may not be sufficient 
to draw any conclusive remark on the dependency of $Q_t$ on the deformation. But still, the value of $Q_t$ 
at this spin matches with that of the value calculated at $\gamma \sim$ $-$75$^\circ$. Rather it lies in between 
the calculated values with $\gamma \sim$ $-$75$^\circ$ and $\gamma \sim$ 30$^\circ$. So, nothing conclusive can be
commented upon. If we compare our result with the results published for $^{142}$Gd, then, we see that their comparison of
B($E$2) values for different minima could not produce a real correspondence between the calculated and observed results for the 
spin range $I$ $\sim$ 16-20 $\hbar$. Though, as they could observe the bands up to a high spin value, therefore, they could 
correlate the calculated B($E$2) values in $\gamma \sim$ $-$75$^\circ$ for the spin range beyond. It is to be mentioned, that, even
their observation was limited up to the state having a spin of 26$\hbar$ and hence the matter was left for open discussion in future attempts. 


It may be noted that the measured $Q_t$ values in $^{140}$Gd exhibit a decreasing trend with spin which indicates the rapid evolution of the shape of the nucleus. This may be indicating the fact that the $\gamma$ rigidity of the core of the nucleus is broken due to the alignment of the valence nucleons. This perhaps makes the shape of the nucleus 
towards the $\gamma$ $\sim$ $-$60$^{\circ}$ at higher spin states giving the possibility for observation of the rotation around classically forbidden longest principal axis. Thus, the present investigation demonstrated that the transition from nuclear rotation about the intermediate axis to the nuclear rotation around the longest principal axis may exist in the $^{140}$Gd nucleus.


\section{Summary}

In summary, the positive parity ground state band structure in $^{140}$Gd has been investigated using heavy-ion induced fusion-evaporation reaction and an array of Compton suppressed Clover detectors. Level lifetimes of the states have been extracted using the Doppler Shift Attenuation Method. Extracted $Q_{t}$ values for the QB I reproduced well within the framework of the pairing independent cranked Nilsson-Strutinsky model calculations. The calculations reveal that the nucleus becomes favourable to rotate about the long axis at the highest observed spin whereas the rotations about the medium axes are observed at intermediate spins. Therefore, the feature of exhibiting the evolution of rotation from intermediate axis to long axis exists in the single nucleus $^{140}$Gd. We hope that this stimulates theoretical and experimental interest in this region, as a great deal of work is required to understand such behavior in $^{140}$Gd and if the phenomenon exists in other nuclei in this mass region.


\begin{center}
$\textbf{ACKNOWLEDGMENTS}$
\end{center}

The untiring effort of the operators at BARC-TIFR pelletron are acknowledged for providing a good beam of $^{35}$Cl. Authors are grateful to Prof. I. Ragnarsson for his constant support regarding CNS calculations and valuable inputs. S. R. would like to acknowledge the financial assistance from the SERB-DST, India under core research grants (File Number: CRG/2020/003370) and the FRPDF grant of Presidency University, Kolkata, India. H. P is grateful for the partial support of the Ramanujan Fellowship Research Grant under SERB-DST (SB/S2/RJN-031/2016), Govt. of India and the Romanian Ministry of Research and Innovation under contract PN 19 06 01 05. S. N. acknowledges the financial support from the SERB-DST, India under RG (File No. : CRG/2021/006671).

\end{document}